\documentclass[journal,a4paper]{IEEEtran}
\usepackage{amsmath,amssymb,amsfonts}
\usepackage{algorithmic}
\usepackage{algorithm}
\usepackage{array}
\usepackage{subfigure}
\usepackage{textcomp}
\usepackage{url}
\usepackage{lipsum}
\usepackage{verbatim}
\usepackage{graphicx}
\usepackage{cite}
\usepackage{color}
\usepackage{xcolor}
\usepackage{multirow}
\usepackage{flushend}
\usepackage{booktabs}
\usepackage {threeparttable}
\usepackage[normalem]{ulem}
\usepackage {soul}

\newcommand{\dB}{\mathrm{dB}}

\renewcommand{\j}{\mathrm{j}}
\newcommand{\R}{\mathrm{R}}
\newcommand{\T}{\mathrm{T}}

\newcommand{\PDP}{\mathrm{PDP}}
\newcommand{\PAP}{\mathrm{PAP}}
\newcommand{\ADPS}{\mathrm{ADPS}}
\newcommand{\DS}{\mathrm{DS}}
\newcommand{\AS}{\mathrm{AS}}

\newcommand{\argmax}{\mathop{\rm arg~max}\limits}

\newcommand{\Vect}[1]{\boldsymbol{#1}}

\newcommand{\Ang}[1]{${#1}^{\circ}$}

\newcommand{\los}{\mathrm{los}}
\renewcommand{\d}{\mathrm{d}}
\renewcommand{\r}{\mathrm{r}}
\newcommand{\BW}{\mathrm{BW}}
\newcommand{\com}{\mathrm{com}}
\newcommand{\ind}{\mathrm{ind}}
\newcommand{\nw}{\mathrm{nw}}
\newcommand{\sw}{\mathrm{sw}}
\newcommand{\PL}{\mathrm{PL}}
\newcommand{\PG}{\mathrm{PG}}
\newcommand{\NA}{---}

\newif\ifMarking
\ifMarking
    \newcommand{\revision}[4]{\colorbox{#1}{#2}{\sout{#3}}{\color{blue}{#4}}}
\else
    \newcommand{\revision}[4]{#4}
\fi

\begin{document}

\title{Quasi-Deterministic Modeling of Sub-THz Band Access Channels in Street Canyon Environments}

\author{Minseok~Kim,~\IEEEmembership{Senior Member,~IEEE,}
        ~Masato~Yomoda,
        ~Minghe~Mao,~\IEEEmembership{Member,~IEEE,}
        ~Nobuaki~Kuno,
        ~Koshiro~Kitao,~\IEEEmembership{Member,~IEEE,} and
        ~Satoshi~Suyama,~\IEEEmembership{Member,~IEEE}
\thanks{M. Kim (e-mail: mskim@eng.niigata-u.ac.jp), M. Yomoda, and M. Mao are with the Graduate School of Science and Technology, Niigata University, Niigata 950-2181, Japan.}
\thanks{N. Kuno, K. Kitao and S. Suyama are with the NTT DOCOMO, INC., Yokosuka-shi, Kanagawa 239-8536, Japan.}
\thanks{This work was partly supported by the Commissioned Research through the National Institute of Information and Communications Technology (NICT) (\#JPJ012368C02701), and the Ministry of Internal Affairs and Communications (MIC)/FORWARD (\#JPMI240410003), Japan.}
}

\maketitle

\begin{abstract}
Sub-terahertz (sub-THz) frequencies (100--300 GHz) are expected to play a key role in beyond-5G and 6G mobile networks. However, their quasi-optical propagation characteristics require new channel models beyond sub-100 GHz extrapolations. This paper presents an extensive double-directional (D-D) channel measurement campaign conducted in an outdoor street-canyon environment at 154 GHz and 300 GHz under both line-of-sight (LoS) and non-line-of-sight (NLoS) conditions using an in-house-developed multi-tone frequency-domain channel sounder. Based on these measurements, clustering with merged datasets across the two frequencies enables comparative analyses that identify both common and distinct multipath clusters, as well as the frequency dependence of cluster-level characteristics. A quasi-deterministic (Q-D) channel model is then proposed, combining deterministic components, such as LoS and single-bounce reflections from side walls, with random components. Large-scale parameters (path loss, delay spread, angular spread, and Rician $K$-factor) are also evaluated. These results provide valuable insights into sub-THz propagation in urban street canyons and contribute toward the development of accurate, channel models for future 6G systems. 
\end{abstract}

\begin{IEEEkeywords}
Sub-terahertz, dual-band channel sounding, outdoor multipath propagation, Quasi-deterministic channel model, street canyon, outdoor hot spot, multipath interaction
\end{IEEEkeywords}


\section{Introduction}
Sixth-generation (6G) mobile communications are expected to achieve ultra-high data rates exceeding $100$~Gbps and ultra-low latency on the order of $100$~$\mu$s, far surpassing the capabilities of current networks. To meet these stringent requirements, research has increasingly focused on the low-terahertz (low-THz) or sub-THz spectrum ($100$--$300$~GHz)\footnote{Throughout this paper, ``sub-THz'' (or ``low-THz'') refers to the $100$--$300$~GHz range; unless otherwise stated, all results and model parameters are derived from measurements at $154$~GHz and $300$~GHz only (i.e., no extrapolation beyond $300$~GHz is performed), while ``THz'' is used only in a general sense to denote frequencies $\geq 0.1$~THz.
}, which offers contiguous channel bandwidths of several tens of GHz. A significant milestone was reached at the World Radiocommunication Conference 2023 (WRC-23), where it was agreed to consider allocating the $102$--$109.5$~GHz, $151.5$--$164$~GHz, $167$--$174.8$~GHz, $209$--$226$~GHz, and $252$--$275$~GHz bands for future International Mobile Telecommunications (IMT) as part of the preliminary agenda for WRC-31 \cite{WRC23Res255}. This regulatory progress, together with growing research in channel measurements, propagation modeling, and early standardization discussions in organizations such as 3GPP \cite{3gppTR38901} and ETSI \cite{etsiTHz2024}, has further strengthened expectations that sub-THz frequencies will become a cornerstone of mobile networks beyond 2030.

Although THz frequencies have so far been applied mainly to point-to-point (P2P) fixed links and backhaul communications, their extension to short-range mobile access is increasingly recognized as promising. According to a 2024 ETSI THz Industry Specification Group (ISG) report \cite{etsiTHz2024}, THz communications could enable a wide range of transformative use cases. However, realizing these applications requires addressing the quasi-optical propagation behavior of THz waves and mitigating their severe propagation losses. THz channels are strongly dominated by line-of-sight (LoS) propagation and experience heavy attenuation in non-line-of-sight (NLoS) conditions due to blockage by both static and dynamic objects. Reliable 100-Gbps mobile access therefore depends on highly directional links with a combined transmitter (Tx) and receiver (Rx) antenna gain exceeding 50~dBi, capable of exploiting both the LoS path and multipath components \cite{Commag_Rikkinen}. Yet, reflection losses, amplified by rough-surface scattering, limit multipath viability as the distance between antennas and reflectors increases \cite{Commag_Kim}. Densifying base station deployments can mitigate coverage gaps but is economically unfeasible. Instead, exploiting multipath through multi-beam management and passive/intelligent reflecting surfaces is essential, though intelligent reconfigurable surface (IRS) technologies still face challenges in control, placement, and cost. A fundamental understanding of sub-THz multipath channels is therefore critical.

\begin{figure}[t]
\centering
\includegraphics[width=0.7\linewidth]{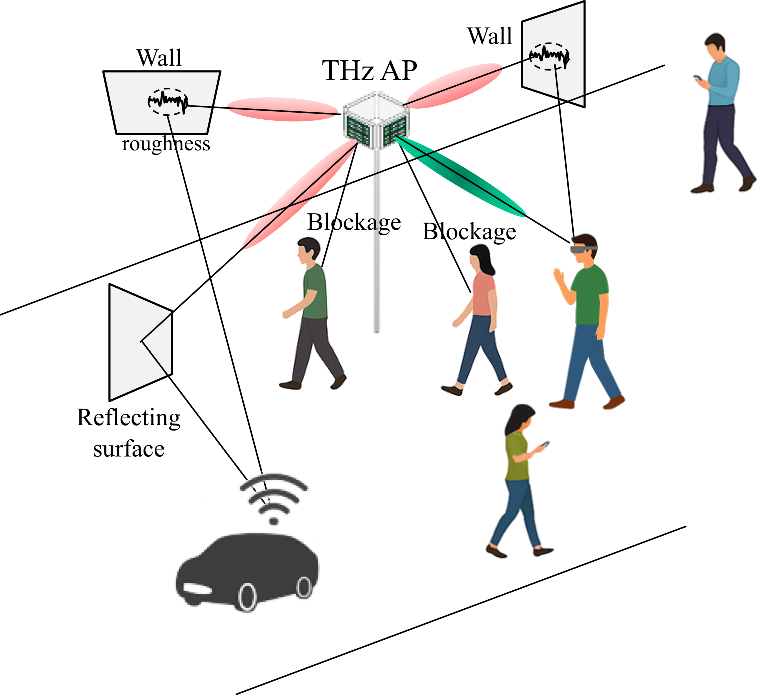}
\caption{THz hot-spot access point in an outdoor street environment. \label{fig:shortcomm}}
\end{figure}

THz hot spots in street environments provide ultra-high-capacity access in dense urban areas, as illustrated in Fig.~\ref{fig:shortcomm}, enabling transformative applications such as immersive augmented reality (AR), virtual reality (VR), mixed reality (MR), connected mobility, and real-time edge computing. Several measurement campaigns have recently investigated sub-THz propagation in urban street canyon environments to support future 6G channel modeling. Angle-resolved measurements at $158$ and $300$~GHz in a UMi street canyon were conducted, highlighting the sparse multipath structure of sub-THz channels \cite{Wilhelm2022}. Similar results were obtained in double-directional (D-D) measurement campaigns at 145~GHz, which revealed dominant LoS and facade reflections with limited angular spread \cite{Shakya2024}. At 140 and 220~GHz, LoS path loss was found to approximate free-space predictions, while NLoS conditions suffered significantly higher attenuation due to building obstructions. Multipath richness was still observed, with notable contributions from facades, vehicles, and street-level scatterers \cite{YangSC2023}. By contrast, 306--371 GHz measurements in a campus street environment revealed much sparser propagation: wall reflections incurred 5--30 dB additional loss, and both delay and angular spreads were substantially smaller than those reported in sub-100 GHz 3GPP street-canyon models. Collectively, these results highlight a clear trend: as frequency increases from 140 to 370 GHz, multipath richness diminishes and channels become increasingly sparse, underscoring the need for new channel models tailored to propagation in this frequency range \cite{WangSC2023}. Additional large-scale efforts by collaborative studies in ETSI and ITU-R have provided datasets \cite{ITU_R_P_1411,ITU_R_P_1238} that are now feeding into early sub-THz channel model development for 3GPP. Collectively, these studies demonstrate that street hot spots are crucial deployment scenarios for THz access, where LoS dominance and a few strong reflections can be effectively exploited for reliable high-capacity connectivity.

As explained above, recent sub-THz channel measurement campaigns have provided valuable insights into propagation in urban street environments; however, several important gaps remain. While path loss and other large-scale characteristics have been reported, detailed multipath analyses are largely missing, leaving open questions about the number and site-specific nature of multipath components that can be effectively exploited. Integrating multiple-input-multiple-output (MIMO) antennas into THz systems enables ultra-narrow directional beamforming to mitigate severe path loss, and multi-stream transmission via multi-beam MIMO can significantly enhance data rates. Consequently, understanding multipath cluster channel characteristics at THz frequencies is essential for accurately assessing the intrinsic channel capacity of an environment and expanding coverage with IRS. From a modeling perspective, current standardized frameworks such as 3GPP TR 38.901 \cite{3gppTR38901} still rely on extrapolations from sub-100 GHz data and assume rich multipath scattering, which does not reflect the sparse structure of sub-THz channels. Consequently, channel model representations that accurately capture frequency-dependent cluster behavior and quasi-deterministic propagation mechanisms in street canyons are still lacking \cite{Commag_Zemen}.

In this context, the present study reports an extensive D-D channel measurement campaign conducted in an outdoor street-canyon environment at 154 and $300$~GHz under both LoS and NLoS conditions using an in-house-developed channel sounder. Based on these measurements, clustering with merged datasets from the two frequency bands is performed, enabling comparative analyses that identify both common and distinct multipath clusters, as well as the frequency dependence of cluster-level characteristics. A quasi-deterministic (Q-D) channel model is then proposed, in which the deterministic components include the LoS path and single-bounce reflections from the two side walls, while the randomness of multipath components is modeled using an approach adapted from corridor scenarios \cite{THzCorridor} owing to their structural similarity (e.g., ``canyon-like'' environments). The deterministic components also incorporate randomness introduced by reflector's surface conditions, represented by a two-state Markov process for presence and absence. Finally, the propagation channel is comprehensively characterized in terms of path loss (PL), modeled using both Close-In (CI) and Floating-Intercept (FI) approaches. Additionally, we analyze large-scale parameters (LSPs), including root-mean-square (RMS) delay spread (DS), azimuth angular spread (AS), and Rician $K$-factor (KF). The main contributions of this work are summarized as: 
\begin{itemize}
\item Extensive dual-band simultaneous D-D channel measurements at 154 and $300$~GHz in a street-canyon environment under both LoS and NLoS conditions using a consistent setup; 
\item Dual-band joint signal processing, including noise filtering for enhanced multipath detection and clustering for comparative analysis across the two frequency bands; 
\item Comprehensive evaluation of large-scale parameters, namely PL, DS, AS, and Rician $K$-factor;
\item Development of a Q-D channel model that extends the corridor-based framework \cite{THzCorridor} to outdoor street canyons by integrating deterministic and random components.
\end{itemize}

The remainder of this paper is organized as follows. Section II outlines the measurement campaign, including scenario descriptions and a brief overview of the channel sounding setup. Section III details the post-processing procedures, such as dual-band joint noise filtering and clustering. Section IV examines the dominant propagation mechanisms by analyzing the trajectories of measured power spectra and validating them through ray-tracing. Section V characterizes the channel in terms of path loss (PL) and large-scale parameters (LSPs). Section VI presents the proposed Q-D channel model and its quantitative validation. Finally, Section VII summarizes the key findings and highlights directions for future research.

\begin{figure*}[t]
    \centering
    \subfigure[Layout. \label{fig:topview}]{\includegraphics[width=0.6\linewidth]{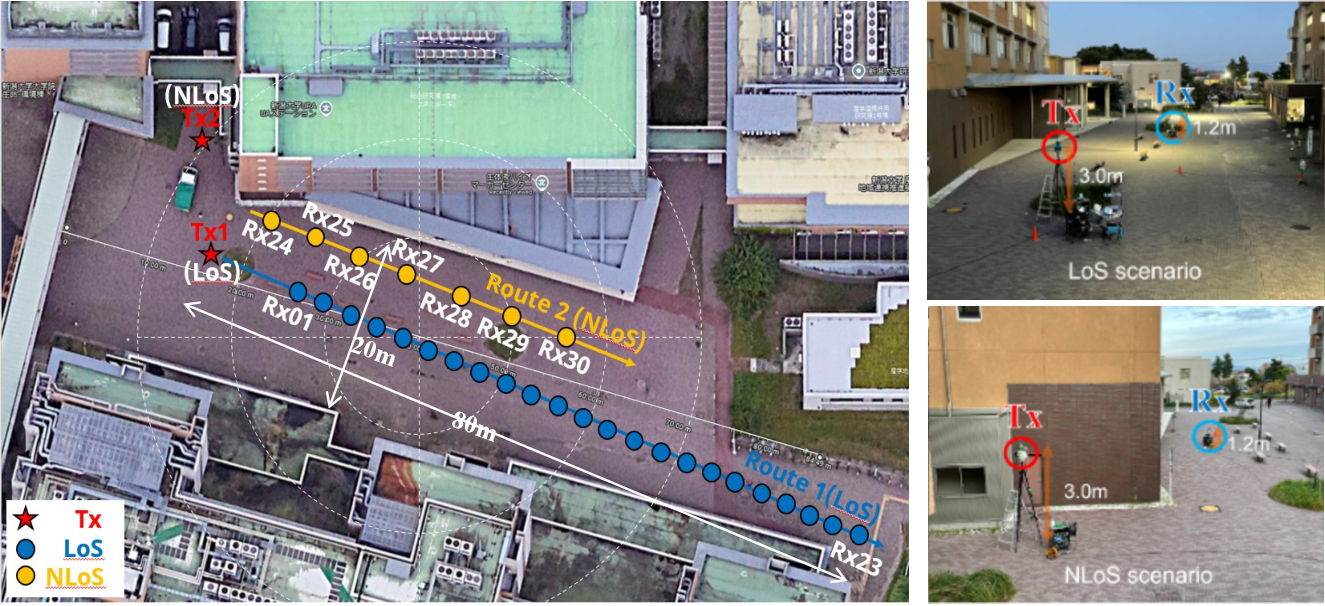}} \quad
    \subfigure[Angle scanning setup. \label{fig:AngleScanning}]{\includegraphics[width=0.33\linewidth]{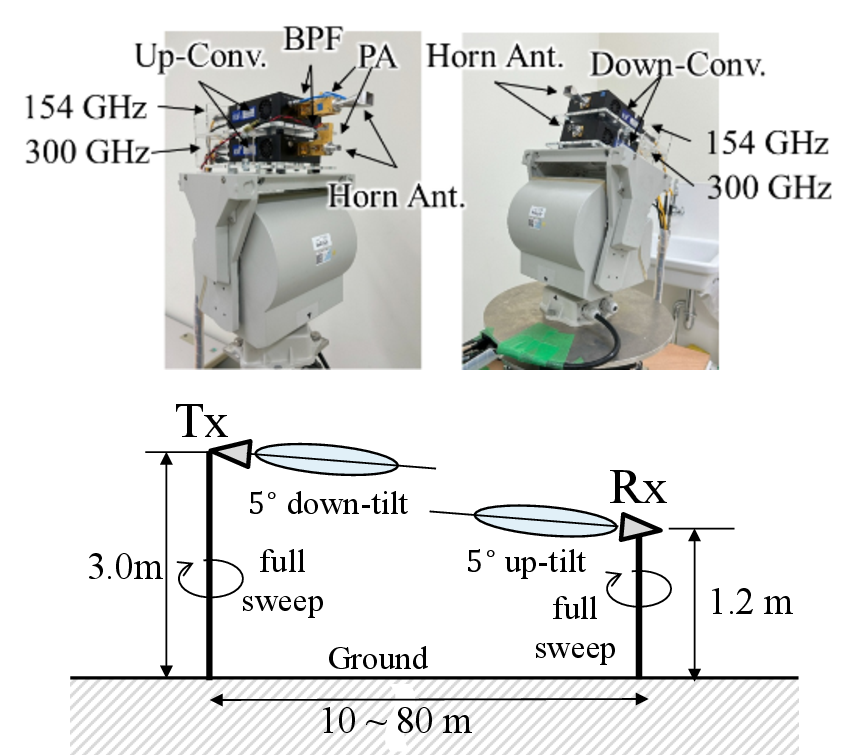}}
    \caption{Measurement setup.}
    \label{fig:On-site}
\end{figure*}

\section{Measurement campaign}

\subsection{Measurement Scenarios}
The D-D channel measurements were conducted in a street canyon with a width of 20--25~m on the Niigata University campus. High-rise buildings (height $> 40$~m) line both sides of the street canyon, as illustrated in Fig.~\ref{fig:topview}. In addition, there are large glass windows on the side walls. During the measurement, the Rx was placed in 30 distinct positions ({\tt Rx1}--{\tt Rx30}) at a height of $1.2$~m from the ground, emulating a user equipment (UE). The Tx was fixed at a height of $3$~m above the ground to emulate a base station (BS) and sequentially positioned at two locations at one end of the street canyon. One location provided a LoS scenario, with {\tt Tx1} corresponding to receiver positions {\tt Rx1}--{\tt Rx23} (distance ranging from 10--80~m). The other location was positioned behind the corner of a building, blocking the LoS path to the receiver and creating an NLoS scenario, with {\tt Tx2} corresponding to receiver positions {\tt Rx24}--{\tt Rx30} (distance ranging from 10--40~m).

Directional scanning measurement was conducted, and the setup is shown in Fig.~\ref{fig:AngleScanning}. On both sides, $26$ dBi gain directional horn antennas with the narrow HPBWs of $\theta_\BW=$\Ang{8} and $\phi_\BW=$\Ang{9} were employed for both bands. The full azimuth (Az) sweep was conducted at both Rx and Tx, whereas the Rx and Tx antenna elevation (El) angles were fixed at \Ang{85} and \Ang{95}, respectively (\Ang{5} tilted). Owing to the simple geometry of the environment, no elevation sweep was conducted \cite{Shakya_ICC,Kim_Access300GHz}; thus, the measurements focused on propagation characteristics in the local horizontal plane. As a result, $40$ angular samples in the azimuth at both Tx and Rx were obtained (1,600 combinations). The antenna gains and radiation pattern effects were removed from the measured data in the post-processing.

\subsection{Channel Sounding}
Measurements were conducted using a $154/300$~GHz dual-band multi-tone frequency-domain channel sounder developed from our earlier work \cite{Kim_Access300GHz}. Since the 300-GHz system has already been described in detail in \cite{Kim_Access300GHz}, this paper highlights the dual-band extensions and their distinctive features. The system parameters are summarized in Table~\ref{parameters}. 

At the Tx side, two intermediate-frequency (IF) signals with a center frequency of 8 GHz are generated using two separate channels of an arbitrary waveform generator (AWG, M8195A, Keysight) operating at a sampling rate of 64 GSa/s. The signal generator (SG) produces a local oscillator (LO) signal at 24.33 GHz, which is fed into two frequency upconverters (WR6.5 CCU and WR3.4 CCU, VDI Corporation). These upconverters multiply the LO frequency by factors of 6 and 12, thereby converting the IF signals to the $154$~GHz and $300$~GHz bands, respectively. Each dual-band signal is then filtered by a band-pass filter (BPF) to suppress image frequencies, amplified by a power amplifier (PA), and transmitted via a horn antenna. At the Rx side, the signals are captured by horn antennas and downconverted using WR6.5 and WR3.4 CCD modules (VDI Corporation) for the 154 and $300$~GHz bands, respectively. The resulting IF signals are amplified by low-noise amplifiers (LNAs) and sampled with a high-speed digitizer (M8131A, Keysight) for storage and subsequent processing. Multi-tone sounding signals were employed, consisting of 2,560 tones over 4 GHz bandwidth at $154$~GHz and 5,120 tones over 8 GHz bandwidth at $300$~GHz. This configuration provided delay resolutions of 250 ps and 125 ps for the 154 and $300$~GHz bands, respectively, while maintaining a common delay span of 640 ns. 

\begin{table}[t]
    \caption{System parameters.}
    \label{parameters}
    \centering
    \begin{tabular}{c|c|c}
        \hline
        Parameters        & Freq I & Freq II        \\ \hline\hline
        Center frequency  & $154$~GHz & $300$~GHz    \\ \hline
        Signal bandwidth  & 4 GHz & 8 GHz             \\ \hline
        No. Subcarriers, $N_\mathrm{F}$  & $2,560$ & $5,120$      \\ \hline
        FFT points, $N_f$ & \multicolumn{2}{c}{$10,240$}  \\ \hline
        Sampling rate     & \multicolumn{2}{c}{\begin{tabular}[c]{@{}l@{}}64 GSa (AWG), 32 GSa (Digitizer)\end{tabular}}  \\ \hline
        Delay resolution  & 250 ps & 125 ps         \\ \hline
        Delay span        & \multicolumn{2}{c}{640 ns}     \\ \hline
        Horn antenna      & \begin{tabular}[c]{@{}c@{}}Gain: 26.4 dBi  \\ HPBW: $9^{\circ}$@Az, \\ $8^{\circ}$@El \end{tabular} & \begin{tabular}[c]{@{}c@{}}Gain: 25.8 dBi  \\ HPBW: $9^{\circ}$@Az, \\ $8^{\circ}$@El \end{tabular}  \\ \hline
        Polarization      & \multicolumn{2}{c}{Vertical}    \\ \hline
    \end{tabular}
\end{table}

To compensate for the effects of system response, a back-to-back (B2B) calibration was performed using $30~\dB$ waveguide attenuators (ATT) prior to the campaign \cite{Kim_Access300GHz}. The antenna gain was de-embedded through over-the-air (OTA) calibration, where the antennas were boresight-aligned to maximize the received power, and the fixed elevation tilt was compensated for during post-processing using the antenna radiation patterns\footnote{The elevation tilt modifies effective antenna gain rather than intrinsic propagation mechanisms. With narrow-beam antennas, this significantly alters coupling to dominant paths. Given the fixed $5^\circ$ tilt, we focused on the local horizontal plane, compensating for antenna pattern effects in post-processing.}. Furthermore, considering the propagation loss associated with the distance between the Tx and Rx, the signal-to-noise ratio (SNR) of the received signal in each band is expected to degrade significantly. This limitation restricts the achievable measurement range. To mitigate this issue, the continuously received sounding symbols are coherently averaged over multiple symbol durations, thereby improving the SNR and extending the effective measurement range.

\section{Post-Processing}
\subsection{Data format}
The data acquired in the measurement is the angle-resolved channel transfer function (CTF), $H(\check{f}, \check{\phi}_\T, \check{\phi}_\R)$, where $\check{f}$, $\check{\phi}_\T$, and $\check{\phi}_\R$ are the frequency bin, and the pointing azimuth angles of departure (AoD) and arrival (AoA), respectively. The channel impulse response (CIR), $h(\check{\tau}, \check{\phi}_\T, \check{\phi}_\R)$, is obtained by taking the inverse Fourier transform of the measured CTF, where $\check{\tau}$ is the delay bin. The omnidirectional power delay spectrum (PDP), the azimuth delayed power spectra (ADPS), and the angular power spectra (APS) can be obtained by summing the relevant spectral bins as
\begin{eqnarray}
\PDP(\check{\tau}) &=& \sum_{\check{\phi}_\T, \check{\phi}_\R} P'(\check{\tau}, \check{\phi}_\T, \check{\phi}_\R), \label{eq:PDP} \\
\ADPS_x(\check{\tau},\check{\phi}_x) &=& \sum_{\check{\phi}_y} P'(\check{\tau}, \check{\phi}_\T, \check{\phi}_\R), \label{ADPSx}\\
\PAP(\check{\phi}_x) &=& \sum_{\check{\tau}, \check{\phi}_y} P'(\check{\tau}, \check{\phi}_\T, \check{\phi}_\R), \label{eq:PAPx}
\end{eqnarray}
where $P^{\prime}(\check{\tau}, \check{\phi}_\T, \check{\phi}_\R)$ is the noise-filtered double-directional angular delay power spectrum (DDADPS) \cite{Kim_Access300GHz,THzCorridor}, $n$ denotes the sample index of the subscripted domain, and $x, y\in\{\T,\R\}$, $y$ is the complementary to $x$. The DDADPS is defined as the expected value of the square of the absolute CIR as
\begin{equation}
P(\check{\tau}, \check{\phi}_\T, \check{\phi}_\R) \triangleq \mathbb{E}|h(\check{\tau}, \check{\phi}_\T, \check{\phi}_\R)|^2.
\label{eq:DDADPS}
\end{equation}
where $\mathbb{E}$ represents the expectation operator.

\subsection{Noise filtering for dual-band datasets}
The measured CIR sample set $h(\check{\tau}, \check{\phi}_\T, \check{\phi}_\R) \in \mathcal{H}_0$ consists of both signal and noise components, namely $\mathcal{H}_0 = \{h_s \in \mathcal{H}_s, h_w \in \mathcal{H}_w\}$, where $\mathcal{H}_s$ and $\mathcal{H}_w$ denote the signal and noise sample sets, respectively. Owing to the sparsity of the sub-THz channel, only a few samples correspond to actual signal components, whereas the majority represent noise. Effective removal of the noise component is therefore essential for reliable channel analysis. The threshold level for noise filtering can be determined from the estimated noise variance, since the noise samples follow a complex Gaussian distribution with zero mean and variance $\mathbb{E}[|h_w|^2] = \sigma_w^2$, implying that the noise power follows an exponential distribution \cite{Dupleich,Kim_CLEAN}.
\begin{figure}[t]
    \centering
    \includegraphics[width=\linewidth]{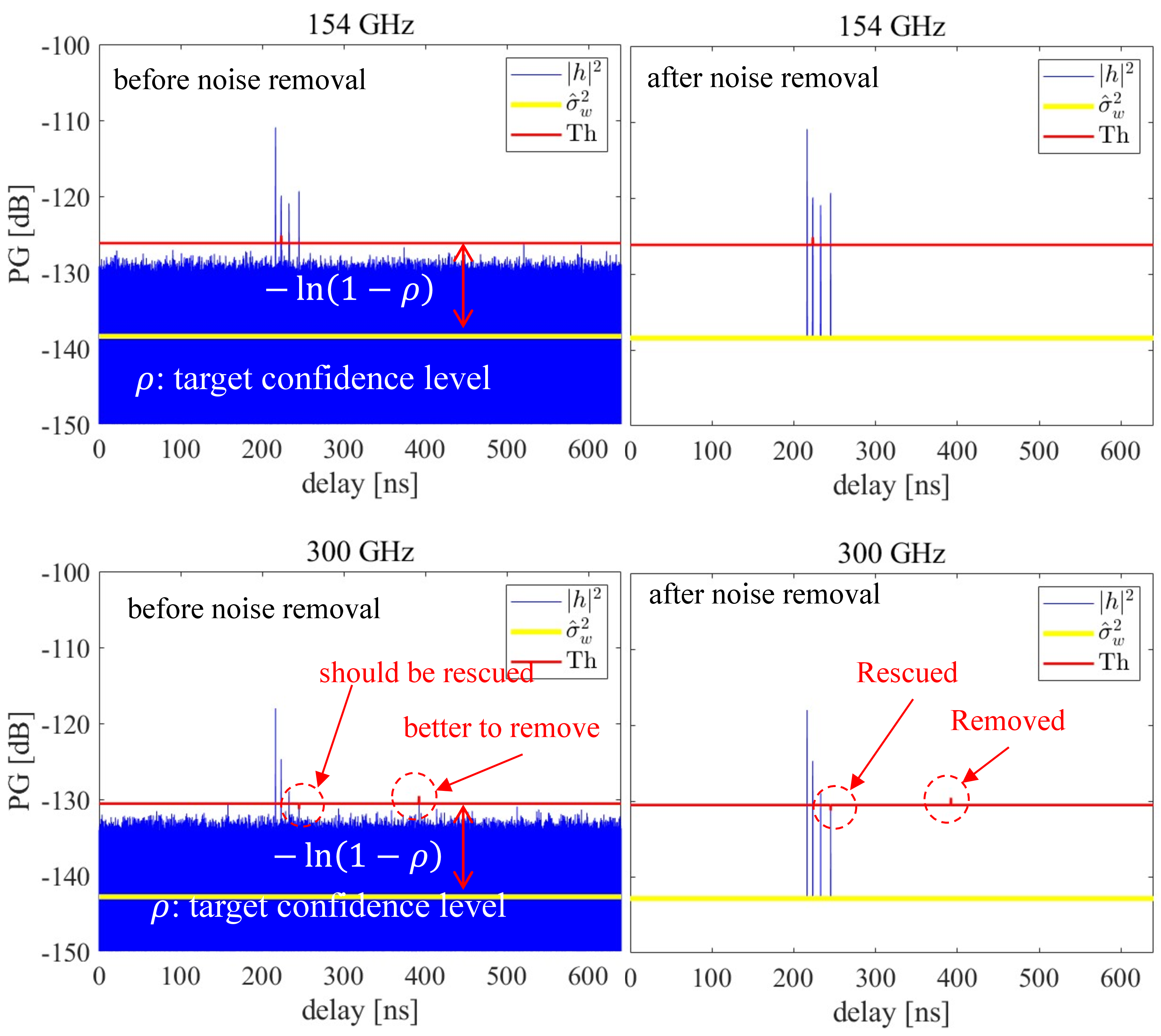}
    \caption{An example of noise filtering applying individual threshold ({\tt Rx19}@$65$~m), where $\xi_\mathrm{FA} = 1$ and $\xi_\mathrm{MD} = 4$ [dB]}.
    \label{fig:noise_filtering}
\end{figure}
\begin{algorithm}[t]
\caption{Determination of individual noise cut-off thresholds for two datasets at bands $f_1$ and $f_2$.}
\label{al:indth}
\begin{algorithmic}[1]
\renewcommand{\algorithmicrequire}{\textbf{Input:}}
\renewcommand{\algorithmicensure}{\textbf{Output:}}
\REQUIRE DDADPS $P_k(\check{\tau}, \check{\phi}_{\T}, \check{\phi}_{\R})$ for bands $k \in \{1, 2\}$.
\ENSURE Adjusted individual thresholds $\zeta_{\ind,k}(\check{\tau})$.

\STATE \textbf{Initialization:}
\STATE Calculate baseline common thresholds $\zeta_{\com,k}$ for each band $k$ using noise statistics.
\STATE Initialize $\zeta_{\ind,k}(\check{\tau}) \gets \zeta_{\com,k}$ for all $\check{\tau}$.
\STATE Detect sets of local peaks $\mathcal{M}_k$ in each band using $\zeta_{\com,k}$.

\STATE \textbf{Inter-band Comparison:}
\STATE Identify unique (uncommon) peak sets $\mathcal{U}_k$ for each band $k$ (peaks present in $\mathcal{M}_k$ but not in $\mathcal{M}_{j}$).
\STATE Set the adaptation margins $\xi_\mathrm{FA}$ and $\xi_\mathrm{MD}$ [dB].

\FOR{each band $i \in \{1, 2\}$}
    \STATE Let $j$ be the complementary band ($j = 3-i$).
    \FOR{each unique peak delay $\check{\tau} \in \mathcal{U}_i$}
        \STATE Obtain spatial angles of the peak in band $i$:
        \STATE \quad $(\check{\phi}_{\T,i}^*, \check{\phi}_{\R,i}^*) = \argmax_{(\check{\phi}_{\T}, \check{\phi}_{\R})} P_i(\check{\tau}, \check{\phi}_{\T}, \check{\phi}_{\R})$
        \STATE Obtain max power and its angles in band $j$ at delay $\check{\tau}$:
        \STATE \quad $P_{j}^{\max} = \max_{(\check{\phi}_{\T}, \check{\phi}_{\R})} P_j(\check{\tau}, \check{\phi}_{\T}, \check{\phi}_{\R})$
        \STATE \quad $(\check{\phi}_{\T,j}^*, \check{\phi}_{\R,j}^*) = \argmax_{(\check{\phi}_{\T}, \check{\phi}_{\R})} P_j(\check{\tau}, \check{\phi}_{\T}, \check{\phi}_{\R})$

        \STATE \textbf{Check Condition:}
        \IF{$P_{j}^{\max} \ge \zeta_{\com,j} - \xi_\mathrm{MD}$ \textbf{AND} $(\check{\phi}_{\T,i}^*, \check{\phi}_{\R,i}^*) = (\check{\phi}_{\T,j}^*, \check{\phi}_{\R,j}^*)$}
            \STATE \COMMENT{\textbf{Rescue in $j$:} Signal exists in margin and angles match (Miss Detection in $j$)}
            \STATE $\zeta_{\ind,j}(\check{\tau}) \gets \zeta_{\com,j} - \xi_\mathrm{MD}$
        \ELSE
            \STATE \COMMENT{\textbf{Remove from $i$:} Power in $j$ is too low or angles mismatch (False Alarm in $i$)}
            \STATE $\zeta_{\ind,i}(\check{\tau}) \gets \zeta_{\com,i} + \xi_\mathrm{FA}$
        \ENDIF
    \ENDFOR
\ENDFOR
\end{algorithmic}
\end{algorithm}
\begin{figure}[t]
\centering
\subfigure[$154$~GHz.\label{Rx03_CL150}]{\includegraphics[width=0.98\linewidth]{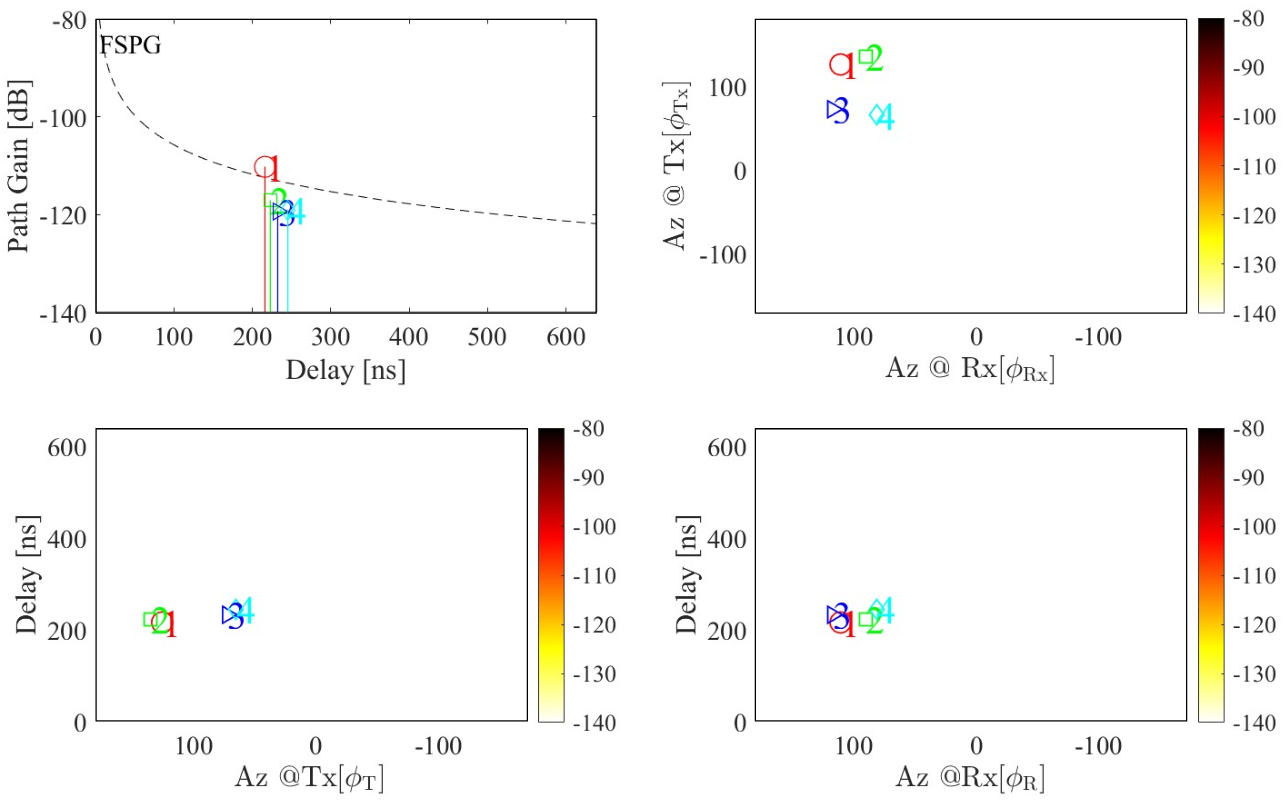}} 
\subfigure[$300$~GHz.\label{Rx03_CL300}]{\includegraphics[width=0.98\linewidth]{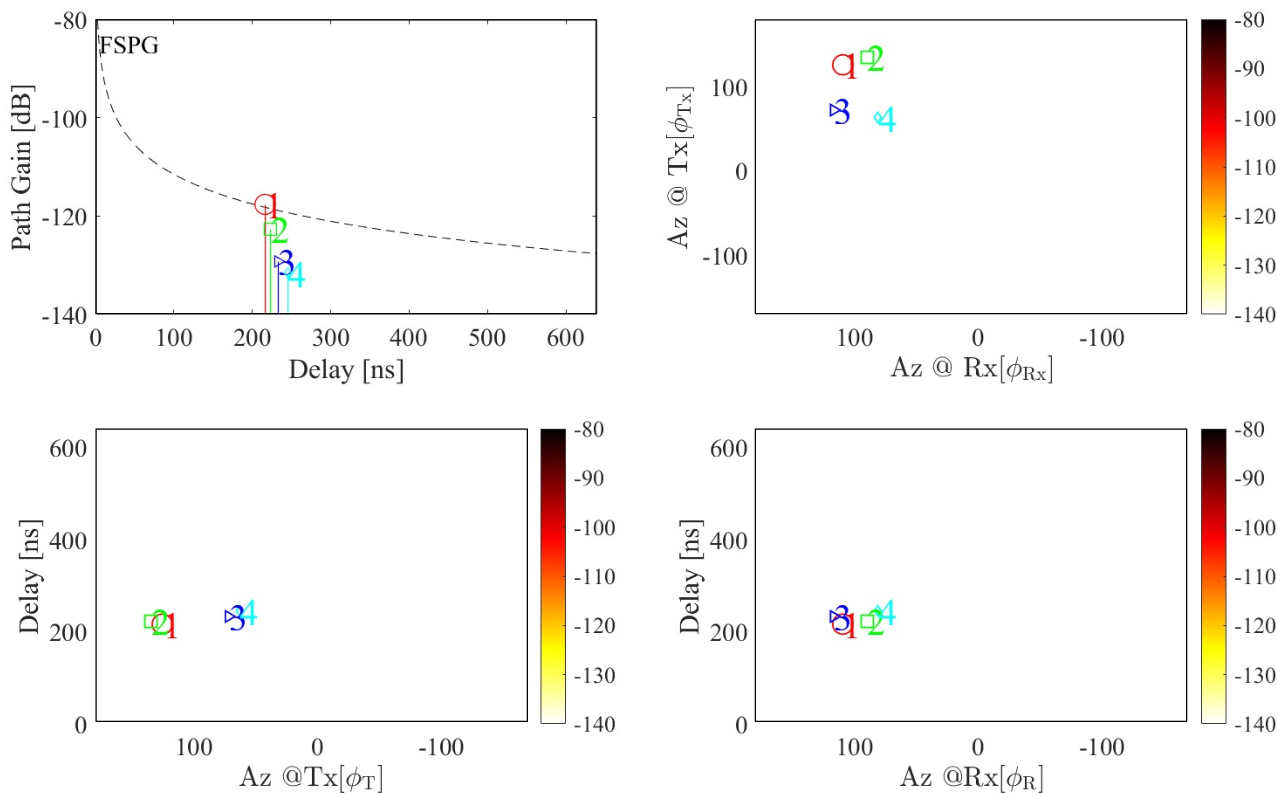}} 
\caption{An example of clustering for merged datasets ({\tt Rx19}@$65$~m, LoS scenario).}
\label{fig:Clustering}
\end{figure}

The noise variance can be estimated as
\begin{eqnarray}
\hat{\sigma}_w^2 &=& \mathbb{E}\left[|h|^2 	\mid h\in\mathcal{H}(1:N_{\mathrm{opt}}) \right],
\label{eq:noisevariance}
\end{eqnarray}
where
\begin{eqnarray}
N_{\mathrm{opt}} &=& \arg \min_{n} \mathbb{E}\left[ \left| F^\mathrm{model}_{\mathcal{H}(1:n)} - F^\mathrm{meas}_{\mathcal{H}} \right|^2 \right],
\end{eqnarray}
$F^\mathrm{meas}_\mathcal{H}$ and $F^\mathrm{model}_{\mathcal{H}(1:n)}$ denote the empirical cumulative distribution function obtained using all samples and the exponential distribution model fitted by the smallest $n$ samples in $\mathcal{H}$ (ascending-sorted version of $\mathcal{H}_0$), respectively. Then, the threshold level is calculated as
\begin{eqnarray}
\zeta_\com = -\hat{\sigma}_w^2 \ln{(1-\rho)},
\label{eq:th_common}
\end{eqnarray}
where $\rho$ is the target confidence level, defined as
\begin{eqnarray}
\rho = 1 - \frac{1}{\nu \cdot |\mathcal{H}|},
\label{eq:rho}
\end{eqnarray}
\revision{green!20}{(R2--5)}{and $\nu$ is a margin designed to effectively suppress false detections, such that one false detection is expected when testing $\nu$ times the total number of samples $|\mathcal{H}|$. Here, $\nu=10$ was chosen.}{where $|\mathcal{H}|$ denotes the total number of CIR samples. The parameter $\nu$ is a margin that controls the expected false-alarm count by setting the per-sample false-alarm probability to $P_{\mathrm{FA}}=1-\rho=1/(\nu|\mathcal{H}|)$. Hence, approximately one false detection is expected among $\nu|\mathcal{H}|$ samples. In this work, we set $\nu=10$.}

The common threshold in \eqref{eq:th_common}, applied to all samples in a single dataset, generally yields satisfactory performance. However, sample sets at two frequencies, even when processed under identical noise-filtering conditions, often produce CIRs that do not align well due to frequency-dependent behavior or false detections. This suggests that noise filtering can be enhanced by cross-checking the two datasets, leveraging the fact that propagation mechanisms remain fundamentally consistent across frequencies. In this approach, the threshold for each frequency is adaptively adjusted along the delay axis by exploiting the physical consistency of multipath propagation across the two bands. This process is designed to mitigate the DR imbalance caused by frequency-dependent losses (e.g., surface roughness). The detailed procedure, summarized in Algorithm~\ref{al:indth}, involves cross-referencing both power levels and angle-domain consistency.

For a unique candidate path detected in band $f_i$ at delay $\check{\tau} \in \mathcal{U}_i$, the corresponding response in the complementary band $f_j$ is examined to distinguish between two scenarios:
\begin{enumerate}
    \item Miss-Detection (Rescue): First, the algorithm checks if a valid signal exists in $f_j$ that was suppressed by the common threshold. If the maximum power in $f_j$ at delay $\check{\tau}$, denoted as $P_j^\mathrm{max}(\check{\tau})$, falls within a specific rescue margin $\xi_\mathrm{MD}$ below the common threshold (i.e., $P_j^\mathrm{max} \ge \zeta_{\com,j} - \xi_\mathrm{MD}$) and the estimated angles in both bands are spatially aligned, the absence of the path in $f_j$ is attributed to a miss-detection. Consequently, the individual threshold for $f_j$ is locally decreased as $\zeta_{\ind,j}(\check{\tau}) \leftarrow \zeta_{\com,j} - \xi_\mathrm{MD}$ to rescue the path.
    
    \item False Alarm (Removal): Conversely, if the component in $f_j$ fails this validation, either because its power is below the rescue margin or because the spatial angles do not match the candidate in $f_i$, the peak in $f_i$ is regarded as a noise artifact (false alarm). In this case, the individual threshold for $f_i$ is locally increased as $\zeta_{\ind,i}(\check{\tau}) \leftarrow \zeta_{\com,i} + \xi_\mathrm{FA}$ to remove this noise component.
\end{enumerate}

This bidirectional validation ensures that weak high-frequency components are preserved only when spatially supported by the lower frequency, while simultaneously preventing the higher-DR band from introducing spurious noise peaks. This strategy effectively mitigates bias favoring the band with a larger dynamic range (154~GHz), particularly when valid components at 300~GHz are significantly attenuated by interaction loss. By enforcing strict angle-domain consistency, the algorithm rescues these attenuated paths using $\xi_\mathrm{MD}$ only if they spatially align with the detection in the reference band; otherwise, the candidate is discarded to preclude false alarms.

Fig.~\ref{fig:noise_filtering} shows an example of clustering ({\tt Rx19}), where all CIR samples are plotted along the delay axis. The adaptation margins are set to $\xi_\mathrm{FA} = 1$~dB and $\xi_\mathrm{MD} = 4$~dB; the former was empirically optimized to balance noise suppression and signal retention, while the latter was determined based on the sensitivity analysis as the saturation point where the recovery of common paths stabilizes. The results demonstrate that multipath components are consistently detected across the two frequencies. It is noted that the resulting dynamic range (DR), defined as the difference between the peak power and the noise floor ($\hat{\sigma}_w^2$), ranges from approximately $20$ to $45$~dB, while the spurious-free dynamic range (SFDR), defined as the difference between the peak power and the threshold level, ranges from $10$ to $31$~dB in the LoS scenario.

\begin{figure*}[t]
\centering
\subfigure[Dominant reflectors (north and south walls). \label{fig:EnvPhoto}]{\includegraphics[width=0.58\linewidth]{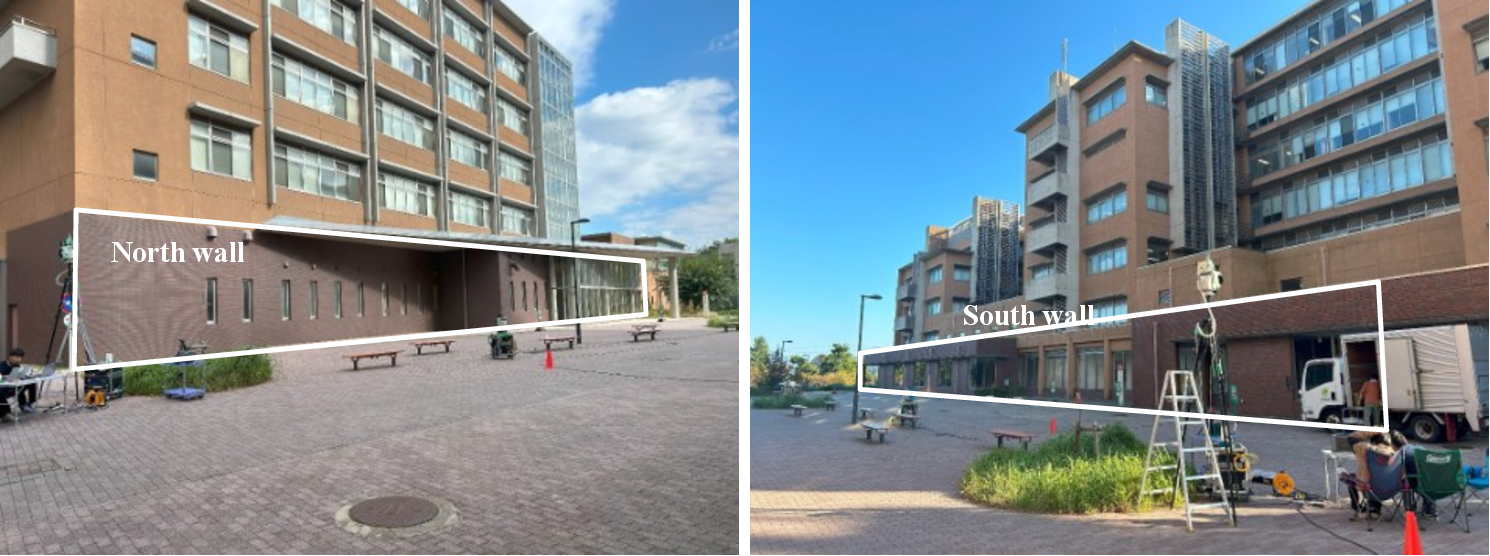}}
\subfigure[Simplified geometrical model (Tx1 for LoS and Tx2 for NLoS). \label{fig:scenarioMap}]{\includegraphics[width=0.36\linewidth]{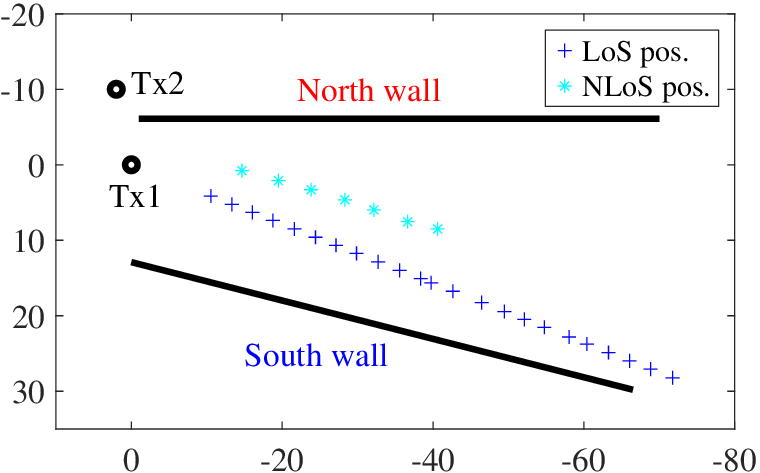}}
\caption{Environment photos and simplified model.\label{fig:EnvPhotoModel}}
\end{figure*}

\begin{figure*}[t]
\centering
\subfigure[LoS scenario.\label{fig:SP_transition_Del_LoS}]{\includegraphics[width=0.45\linewidth]{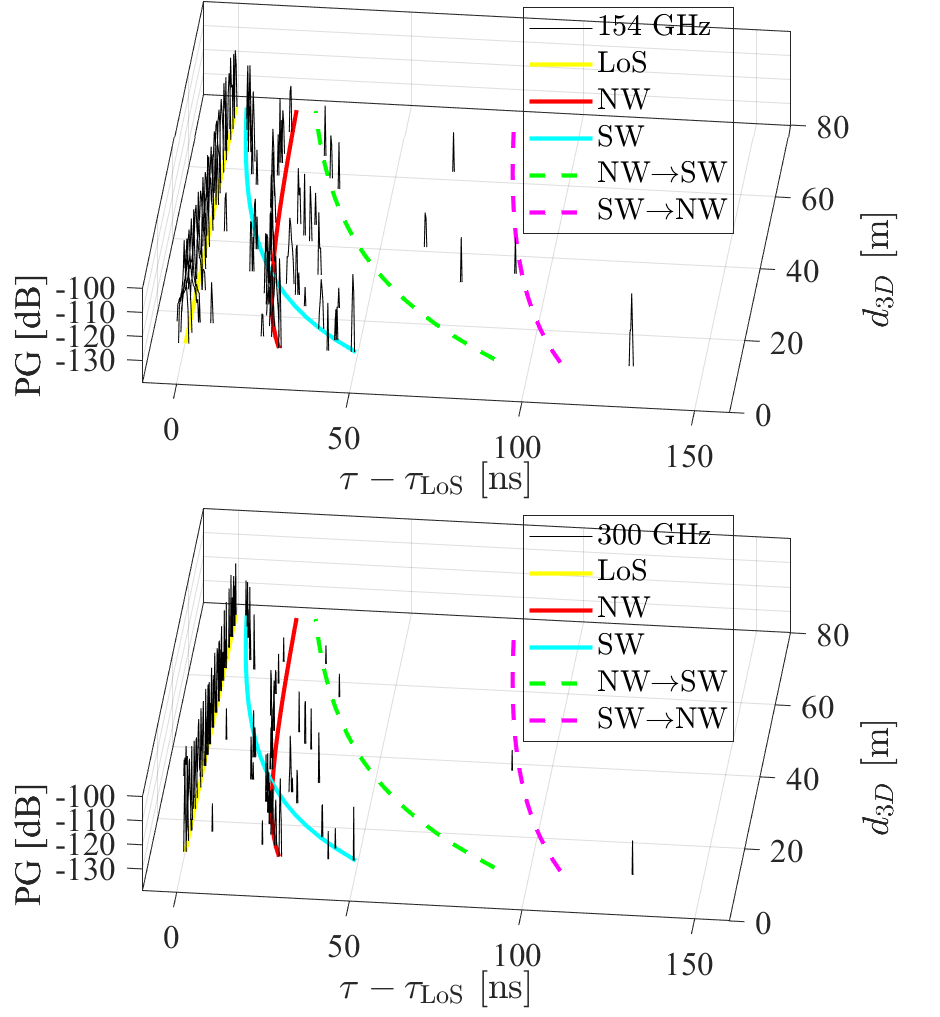}}
\subfigure[NLoS scenario.\label{fig:SP_transition_Del_NLoS}]{\includegraphics[width=0.45\linewidth]{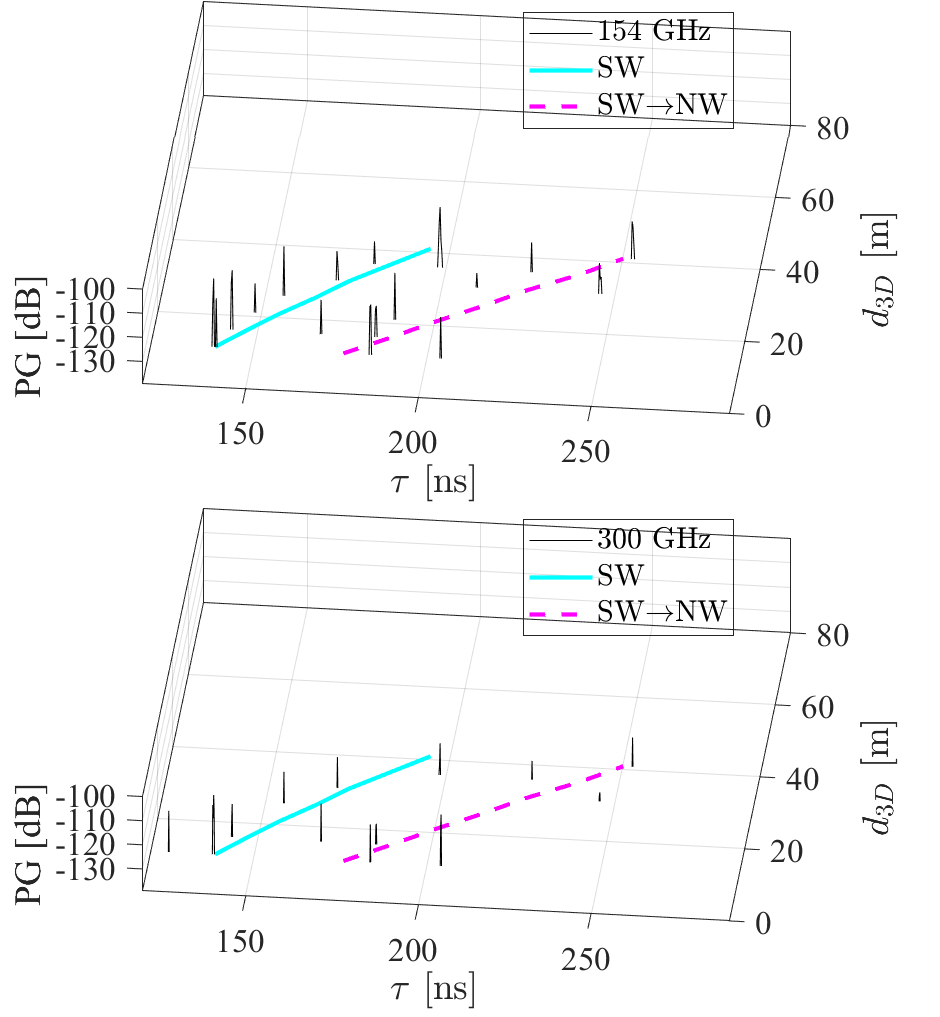}}
\caption{PDP variations along the Rx trajectory, where left: $154$~GHz, right: $300$~GHz. \label{fig:SP_transition_Del}}
\end{figure*}

\begin{figure*}[t]
\centering
\subfigure[Relative delay.\label{fig:CL_transition_Del}]{\includegraphics[width=0.328\linewidth]{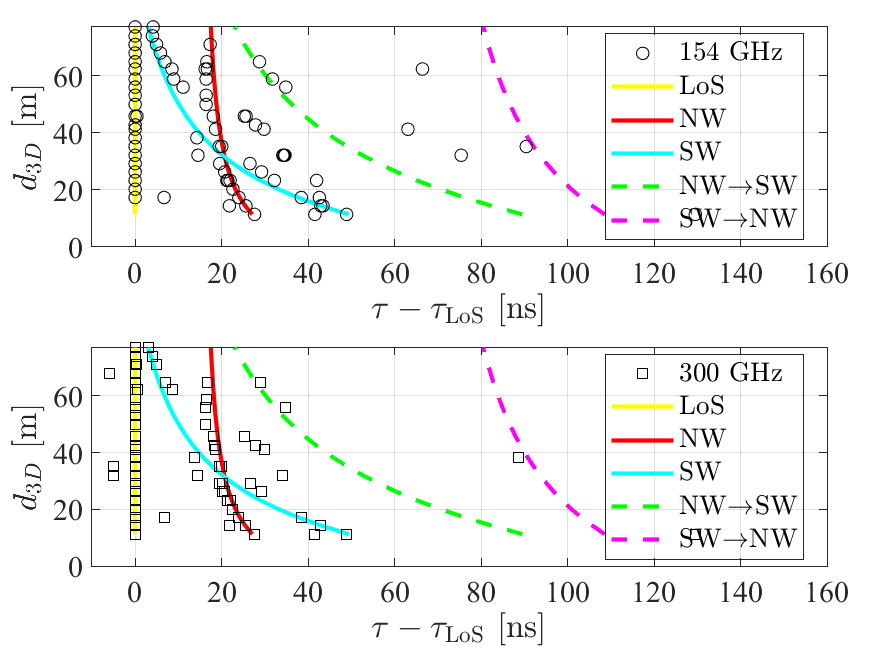}} 
\subfigure[Relative AoD.\label{fig:CL_transition_AoD}]{\includegraphics[width=0.328\linewidth]{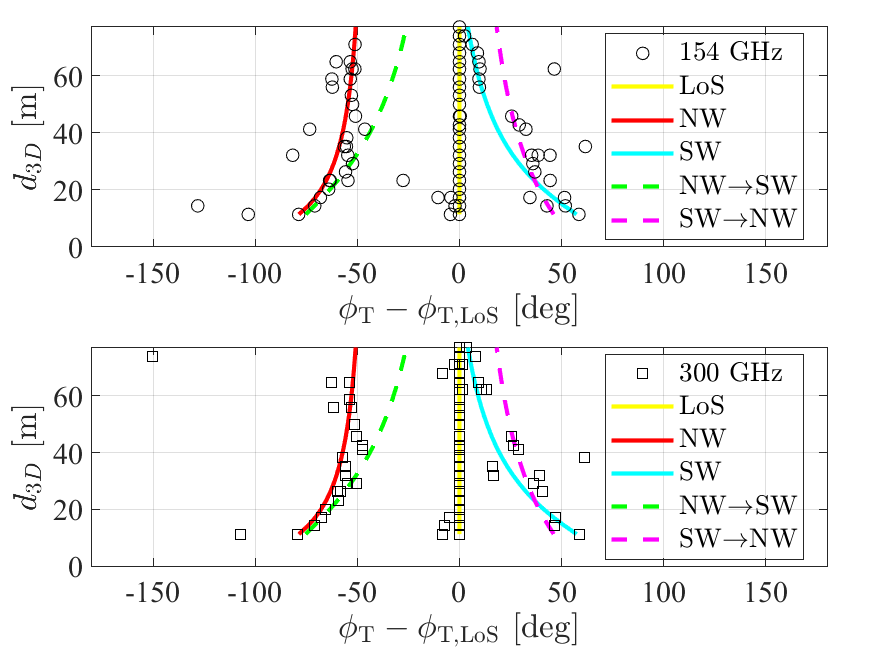}} 
\subfigure[Relative AoA.\label{fig:CL_transition_AoA}]{\includegraphics[width=0.328\linewidth]{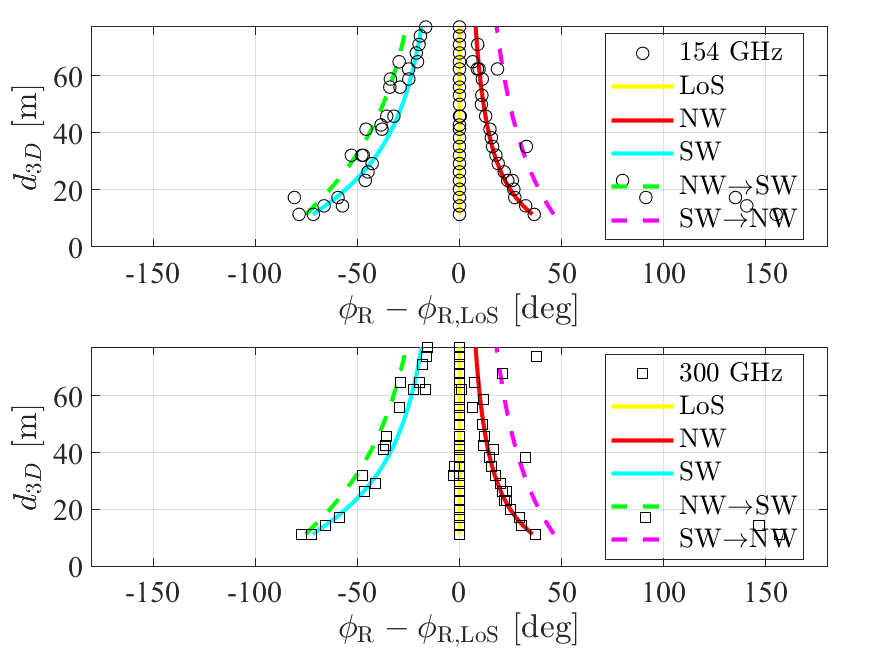}} 
\caption{MPC distributions along the Rx trajectory (LoS scenario).}
\label{fig:ClusterTrajectory}
\end{figure*}

\subsection{Clustering}
The multipath components (MPCs) were extracted using the Sub-grid CLEAN algorithm \cite{Kim_CLEAN}, which estimates the parameters of each MPC by sequentially subtracting a replica from the DDADPS in descending order of power. Each replica is generated using the antenna radiation pattern and the signal autocorrelation function. Clustering was applied to the extracted MPCs to group them into clusters based on similar angle and delay parameters. The clustering process was performed on the merged datasets of MPCs obtained at both frequencies under a consistent setup, in order to assign identical cluster indices across the two frequency sample sets \cite{TVT_Tsukada}. To account for the increased propagation loss due to the frequency difference, the power of each $300$~GHz MPC was scaled up by $20\log_{10}\tfrac{300}{154} = 5.792$~dB prior to clustering. After clustering, the clustered MPCs were separated back into their respective frequency datasets, and the power of each $300$~GHz MPC was restored to its original value. This approach enabled the identification of both common clusters and uncommon (frequency-specific) clusters. The $K$-PowerMeans clustering algorithm was employed, with the number of clusters, $K$, determined manually by visual inspection to preserve the physical interpretability of the results. Fig.~\ref{fig:Clustering} presents an example clustering result for {\tt Rx19}, showing four clusters that are common across both frequency bands.

\begin{figure*}[t]
\centering
\subfigure[Distributions of cluster numbers.\label{fig:CL_number}]{\includegraphics[width=0.32\linewidth]{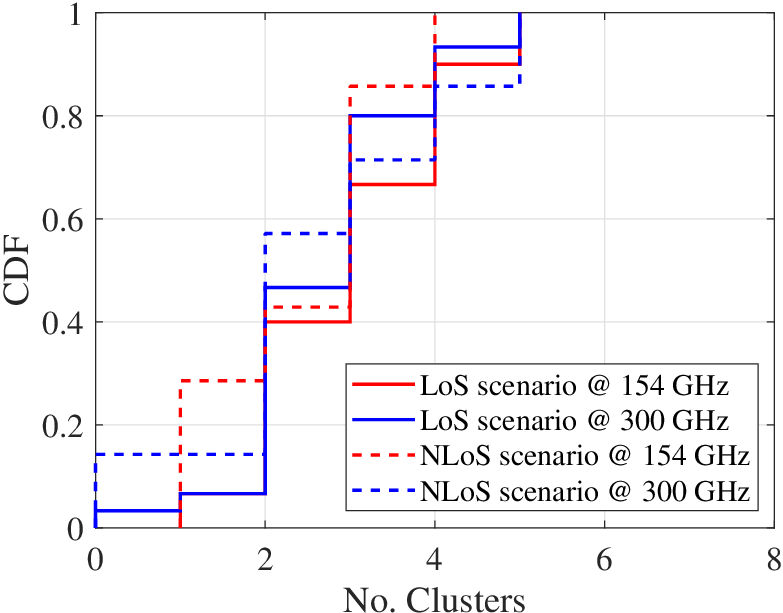}} 
\subfigure[Distributions of LoS cluster power and sum of all NLoS clusters powers (LoS scenario).\label{fig:CLpowerCDF}]{\includegraphics[width=0.334\linewidth]{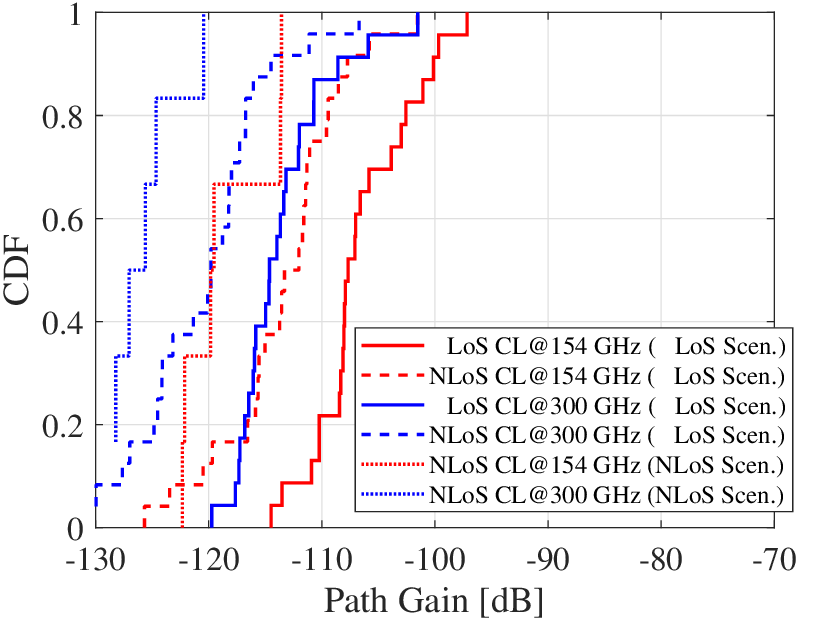}} 
\subfigure[Relative power distribution of NLoS common clusters.\label{fig:CL_Power_FreqDiff}]{\includegraphics[width=0.32\linewidth]{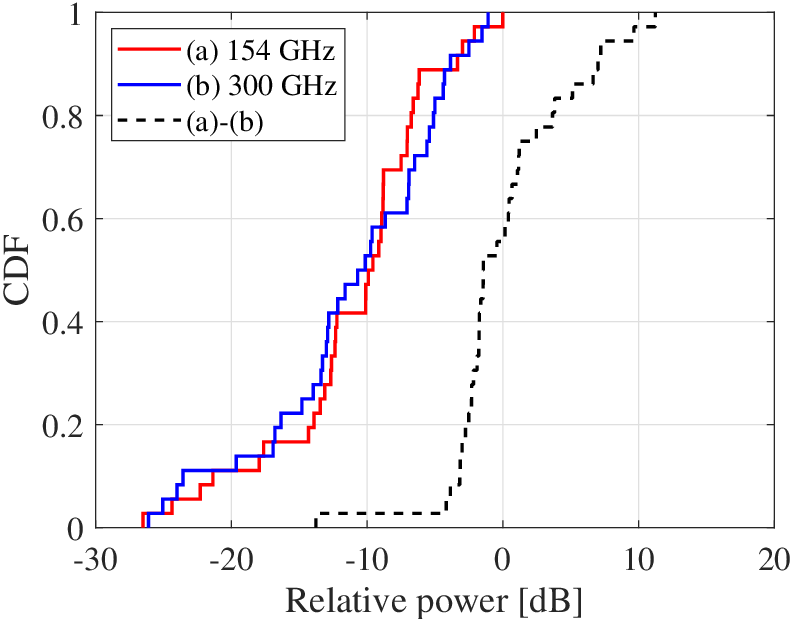}} 
\caption{Cluster characteristics.}
\label{fig:CL_Character}
\end{figure*}

\section{Multipath Cluster Behavior}
\subsection{Dominant Scattering Process}
Fig.~\ref{fig:EnvPhoto} shows photos of the surrounding buildings located to the north and south. The building walls, composed of glass windows/doors and block facades, serve as the primary reflectors that generate the strong single-bounce (SB) paths observed in the measurement campaign. It is noted that the north wall is not perfectly parallel to the street, and there are no trees in front of the walls. Fig.~\ref{fig:scenarioMap} illustrates a simplified two-dimensional (2D) geometrical model of the LoS scenario, which is used for ray-tracing calculations. The material constants for both walls, primarily composed of concrete, are characterized by the following relative complex permittivities: $\varepsilon_c = 6.08 - \j0.153$ at $154$~GHz and $\varepsilon_c = 5.24 - \j0.38$ at $300$~GHz \cite{concrete}. 

Figs.~\ref{fig:SP_transition_Del} depicts the spatial variation of the PDPs along the Rx trajectory for both scenarios. Specifically, Fig.~\ref{fig:SP_transition_Del_LoS} shows the relative PDPs, where the delay is normalized to the LoS delay, while Fig.~\ref{fig:SP_transition_Del_NLoS} presents the normal PDPs as a function of the absolute delay. At both frequencies, the propagation characteristics exhibit pronounced sparsity and simplicity. In the LoS scenario, as shown in Fig.~\ref{fig:SP_transition_Del_LoS}, two major scattering processes that vary consecutively along the Rx trajectory are identified as SB reflections from the north wall (NW) and south wall (SW), with their trajectories validated using the image method ray-tracing applied to the simplified model in Fig.~\ref{fig:scenarioMap}. Aside from the three main components, such as the LoS path and the two SB reflections, only a few additional components are observed, mainly originating from double-bounce (DB) reflections (NW$\rightarrow$SW and SW$\rightarrow$NW) and other unidentified interactions with random objects. This phenomenon becomes more pronounced at $300$~GHz. In contrast, in the NLoS scenario, as shown in Fig.~\ref{fig:SP_transition_Del_NLoS}, two dominant propagation mechanisms that evolve consecutively along the Rx trajectory are approximately identified as the SB reflection from the SW and the DB reflection resulting from the SW$\rightarrow$NW interaction. Note that although the SW$\rightarrow$NW DB interaction can dominate in the NLoS example in Fig.~\ref{fig:SP_transition_Del_NLoS}, this work develops a Q-D model for the LoS scenario because the NLoS dataset is limited. The behavior of the extracted MPCs is illustrated in Figs.~\ref{fig:CL_transition_Del}, \ref{fig:CL_transition_AoD}, and \ref{fig:CL_transition_AoA}, which present the variations in MPC distribution across the delay, AoD, and AoA domains, respectively, at both frequencies. In the LoS scenario, the SB reflections from the NW and SW dominate, whereas DB reflections are rarely observed, a trend that was further confirmed by the clustering results.

\begin{figure*}[t]
\centering
\subfigure[LoS.\label{LoS_PL}]{\includegraphics[width=0.4\linewidth]{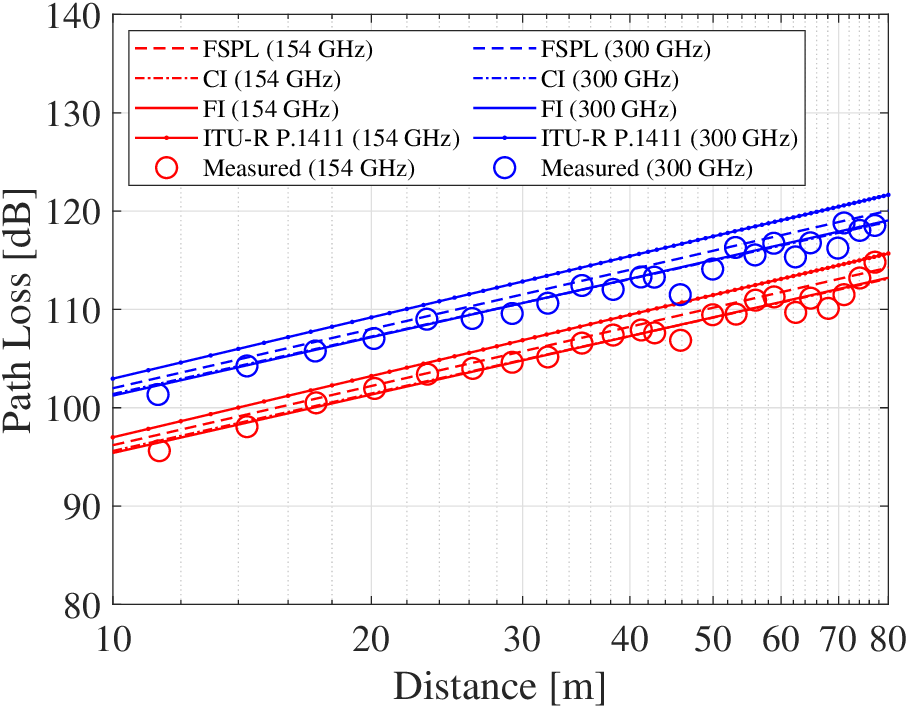}} \qquad
\subfigure[NLoS.\label{NLoS_PL}]{\includegraphics[width=0.4\linewidth]{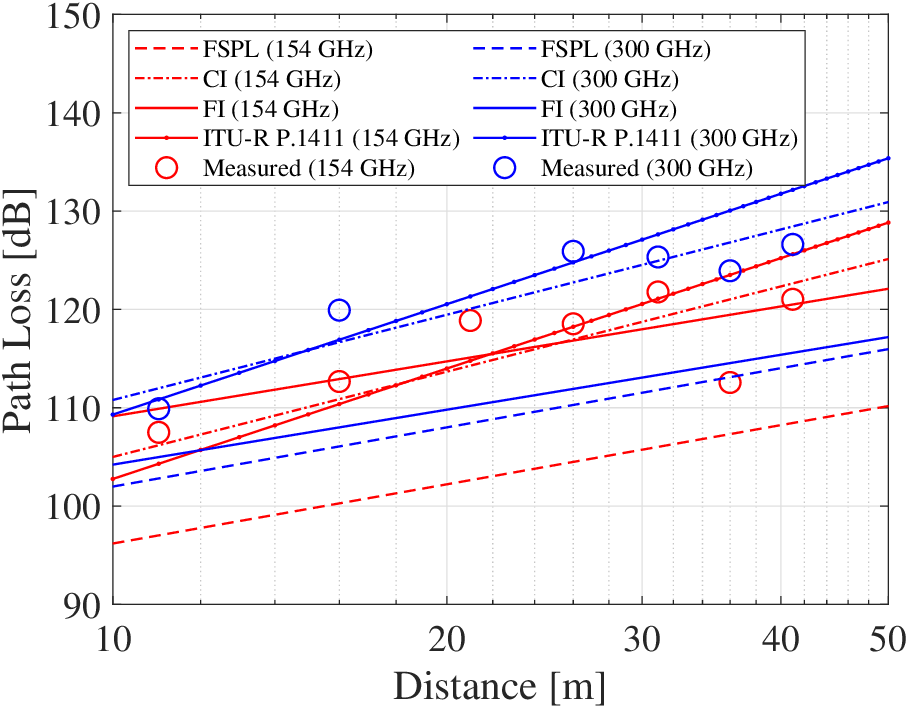}} 
\caption{PL characteristics. \revision{green!20}{(R2--1)}{}{The ABG parameters used for the ITU-R P.1411-13 comparison curves are as follows: LoS: $\alpha = 2.07$, $\beta = 31.23$, $\gamma = 2.06$; NLoS: $\alpha = 3.73$, $\beta = 16.02$, $\gamma = 2.26$~\cite{ITU_R_P_1411}.}}
\label{fig:omniPL}
\end{figure*}

\subsection{Cluster Characteristics}
Fig.~\ref{fig:CL_Character} illustrates the cluster characteristics through the cumulative distribution function (CDF). Fig.~\ref{fig:CL_number} depicts the distribution of the number of clusters, showing that the $154$~GHz band exhibits slightly more clusters than the $300$~GHz band. This difference can be attributed to the higher interaction losses at $300$~GHz. It is also observed that the number of clusters in the NLoS scenario is slightly smaller than that in the LoS scenario.

Fig.~\ref{fig:CLpowerCDF} shows the path gain (PG) distribution of the clusters for each frequency band. The solid lines represent the LoS cluster PG across all Rx points in the LoS scenario, while the dashed lines denote the total PG of the NLoS clusters. 
For the LoS clusters, the average PG values are $-107.67$~dB at $154$~GHz and $-114.63$~dB at $300$~GHz, corresponding to a power difference of $6.96$~dB. This difference is $1.17$~dB larger than the additional FSPL of $5.792$~dB, indicating that the $300$~GHz band experiences slightly greater degradation.
However, for the NLoS clusters, the average PG values are $-113.29$~dB at $154$~GHz and $-119.80$~dB at $300$~GHz, yielding a power difference of $6.51$~dB, which is $0.72$~dB larger than the additional FSPL.

As explained above, the PGs of the NLoS clusters at $300$~GHz exhibit slightly greater degradation compared to those at $154$~GHz. To further examine this, Fig.~\ref{fig:CL_Power_FreqDiff} presents a comparative analysis based on the relative powers of common clusters, which helps mitigate practical calibration issues such as inaccurate antenna gain. The results show that the statistical distribution of NLoS cluster power relative to the LoS cluster power is quite similar for both frequency bands. Moreover, the dashed black curve, representing the difference between the two frequencies, remains roughly balanced, indicating no significant disparity between the $154$~GHz and $300$~GHz bands, despite the wide distribution range from $-10$ to $10$~dB. It is also observed, however, that degradation is more pronounced at $300$~GHz when the relative path gain difference is positive.

\section{Path Loss and LSP Characterization}
Regardless of the specific channel model, channel parameters including PL and LSPs are essential for applying standard channel models such as 3GPP TR 38.901 \cite{3gppTR38901}. In this section, the measured channels are characterized in terms of PL, DS, AS, and Rician $K$-factor at both frequencies and in both scenarios.

\subsection{Path Loss}
The omnidirectional PL, $\mathrm{PL}_{\mathrm{omni}}$ is calculated as 
\begin{equation}
\mathrm{PL}_{\mathrm{omni}} = -\Bigg(10\log_{10}\bigg(\sum_{\forall} {P^{\prime}(\check{\tau}, \check{\phi}_\T, \check{\phi}_\R)} \bigg)-G_\mathrm{A} \Bigg) [\dB].
\label{eq.PL}
\end{equation} 
where $G_\mathrm{A}$ denotes the effect due to incremental angular steps below the HPBW and non-orthogonal radiation patterns of the used antennas which results in an overlap of the antenna beam which leads to an increase in the received power by $G_\mathrm{A}$ as described in Recommendation ITU-R P.1407-8 \cite{ITU_R_P_1407}. In this study, $G_\mathrm{A}$ was analytically calculated based on the measured antenna patterns \revision{green!20}{(R2--1)}{}{\cite{OmniSynth}}. Note that the large measurement bandwidth allows almost all multipath components to be well separated in the delay domain. Consequently, the incoherent power summation in \eqref{eq.PL} yields a fading-free PL, irrespective of the antenna beamwidth. The model fitting parameters are evaluated in the following subsection.

To investigate the characteristics of the dual-band channels under different scenarios, two popular channel models: the close-in (CI) free-space reference distance model \cite{cimodel} and the floating intercept (FI) model \cite{fimodel}, are used for data fitting. 
\begin{itemize}
    \item CI model:
    \begin{equation}
    \label{CI}
        \PL_\mathrm{CI}(d)~\mathrm{[dB]}= 10n\log_{10}(d) + 20\log_{10}\frac{4\pi f_\mathrm{c}}{c} +  \chi,
    \end{equation}
    where, $d$ is the Tx-Rx separation distance, $n$ is the path loss exponent (PLE),  $f_\mathrm{c}$ is the center frequency, and $c$ is the speed of light in $\mathrm{m}/\mathrm{s}$. $\chi$ denotes the log-normal shadow-fading term with zero mean and standard deviation $\sigma_\chi$.
    
    \item FI model:
    \begin{equation}
    \label{FI}
        \PL_\mathrm{FI}(d)~\mathrm{[dB]} = 10\alpha\log_{10}(d) + \beta +\chi,
    \end{equation}
    where, $\alpha$ and $\beta$ are the slope and floating intercept respectively. 
\end{itemize}

\begin{table}[t]
    \centering
    \caption{PL model parameters.}
    \label{PLparameters}
    \begin{tabular}{c c c c}
        \toprule
        \multirow{2}{*}{Scen.} & \multirow{2}{*}{Freq.}  
        & CI Model & FI Model \\
        & & $(n,\,\sigma_\chi)$ & $(\alpha,\,\beta,\,\sigma_\chi)$ \\ 
        \midrule
       
        \multirow{2}{*}{LoS} 
        &$154$~GHz& $(1.94, 0.85)$  & $(1.97$,\;\revision{green!20}{(R2--1)}{$75.64$}{$75.67$},\;$0.85)$ \\ 
        &$300$~GHz& $($\revision{green!20}{(R2--1)}{$1.91$}{$1.92$}, $0.82)$  & $(1.95$,\;\revision{green!20}{(R2--1)}{$81.47$}{$81.50$},$\;0.81)$ \\ \hline
        \multirow{2}{*}{NLoS$^*$}
        &$154$~GHz& $(2.88, 4.26)$  & $(1.86,\;90.57,\;3.72)$ \\ 
        &$300$~GHz& $(2.88, 2.71)$  & $(2.62,\;85.65,\;2.65)$ \\ 
        \bottomrule
        \multicolumn{4}{l}{\scriptsize $^{\mathrm{*}}$Preliminary values due to limited sample size.} \\
    \end{tabular}
\end{table}

\begin{figure*}[t]
\centering
\subfigure[DS.\label{fig:SP_DS}]{\includegraphics[width=0.4\linewidth]{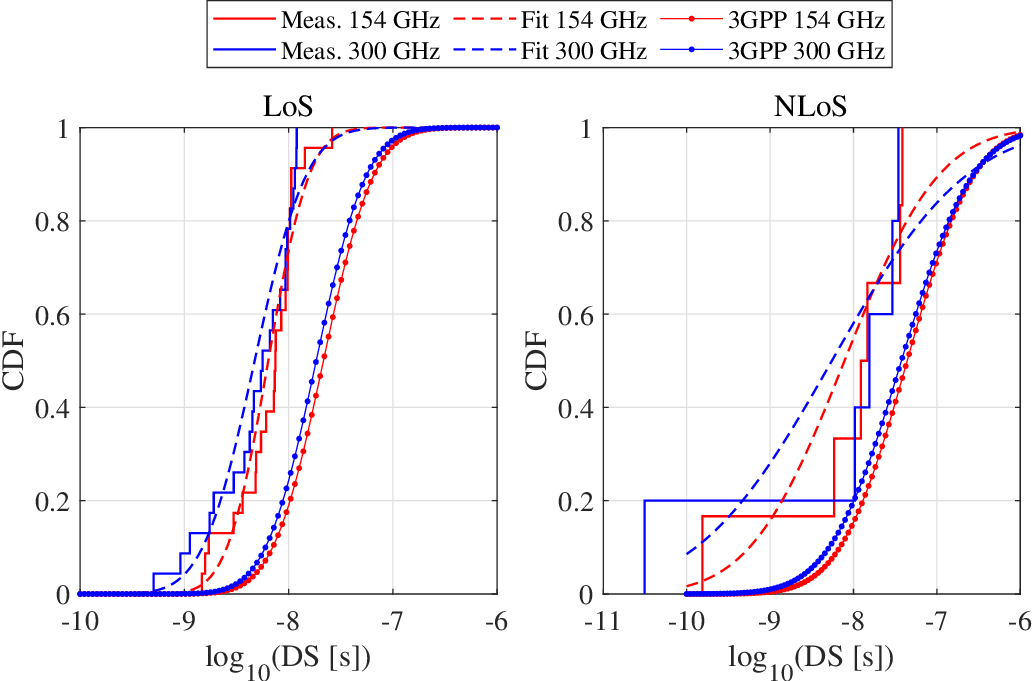}}  \qquad
\subfigure[ASD.\label{fig:SP_ASD}]{\includegraphics[width=0.4\linewidth]{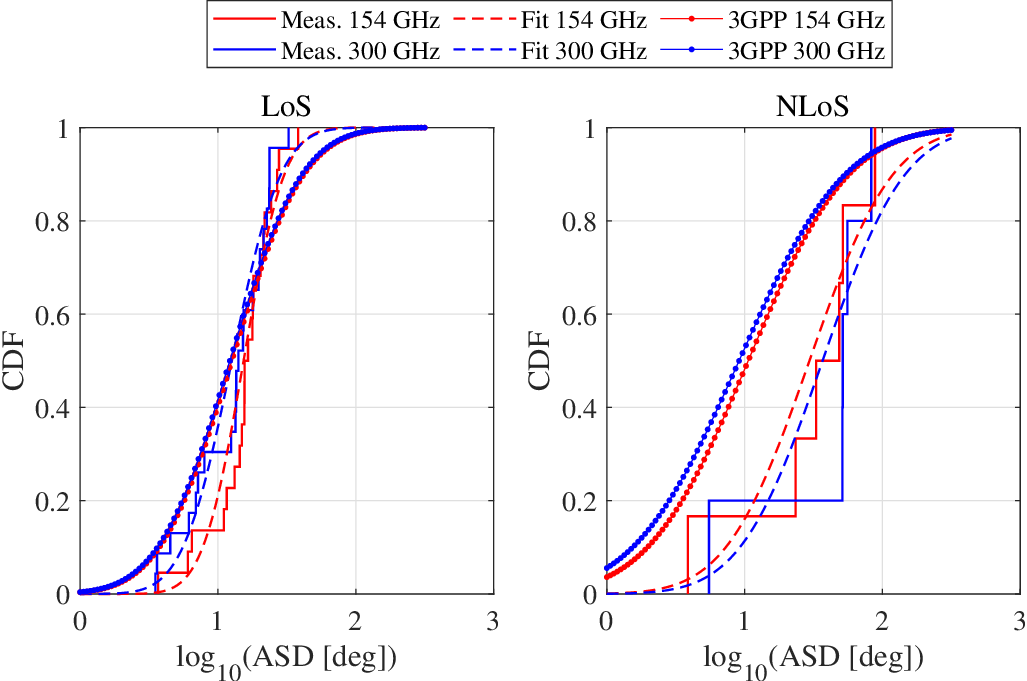}} 
\subfigure[ASA.\label{fig:SP_ASA}]{\includegraphics[width=0.4\linewidth]{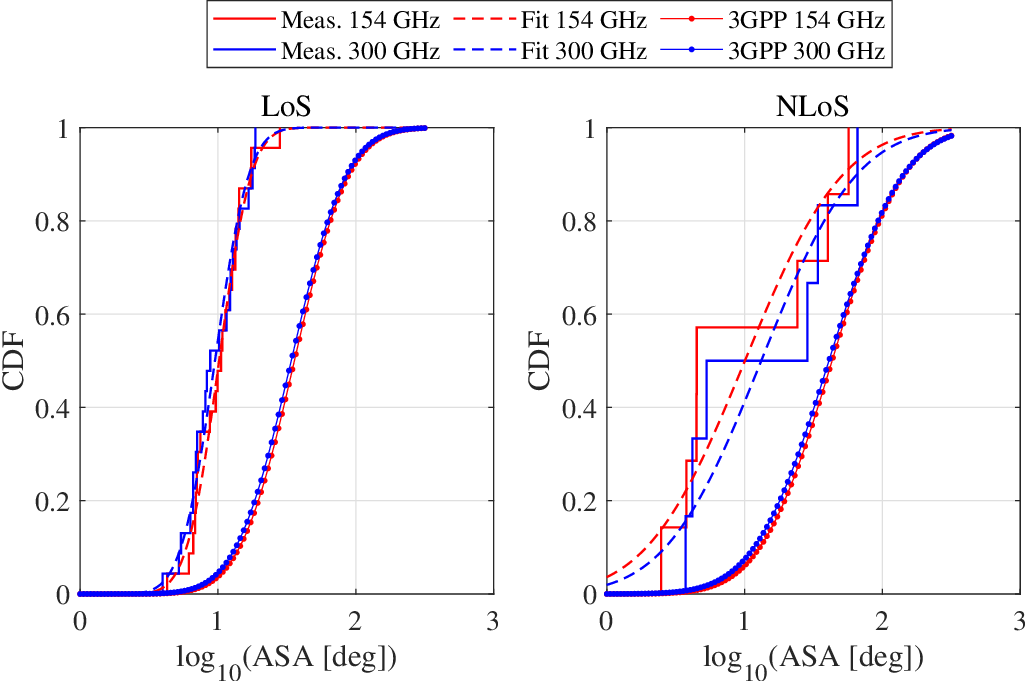}}  \qquad
\subfigure[Rician $K$-factor.\label{fig:CL_Kfactor}]{\includegraphics[width=0.4\linewidth]{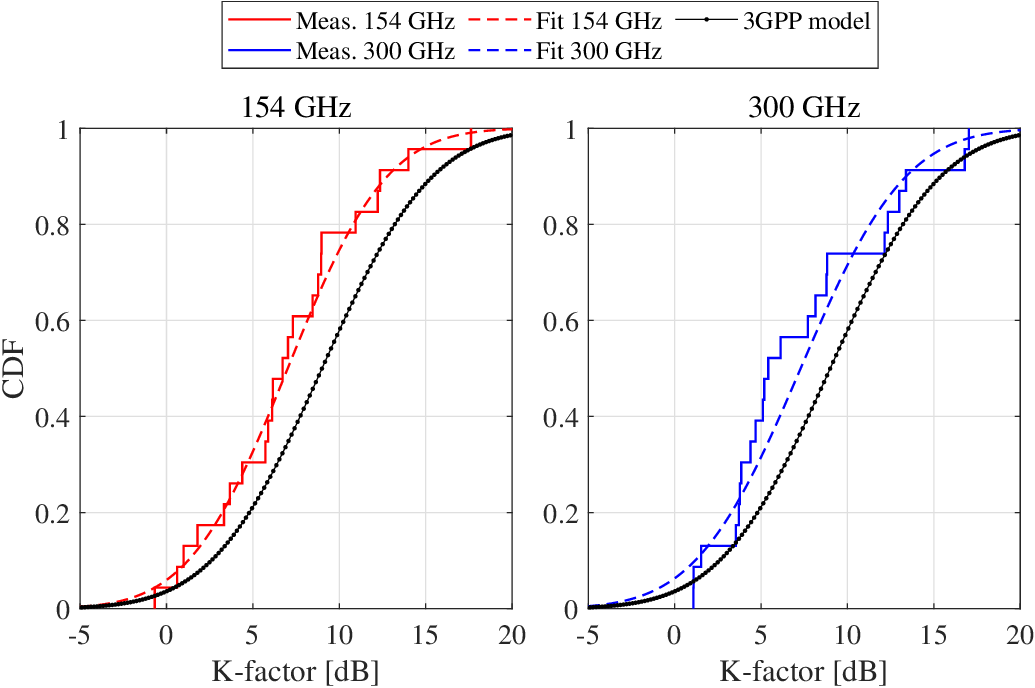}} 
\caption{Distributions of various LSPs, including DS, ASD, ASA, and Rician $K$-factor. Note that the 3GPP model is specified only up to $100$~GHz; thus, the values are extrapolated.}
\label{fig:LSP_dist}
\end{figure*}

The results of fitting the omnidirectional PLs for both LoS and NLoS scenarios using the aforementioned models are shown in Fig.~\ref{fig:omniPL}, and the corresponding model parameters are summarized in Table~\ref{PLparameters}. For comparison, the latest ITU-R P.1411-13 site-general model (ABG model), which supports frequencies up to 300~GHz, is also plotted as a reference. 
\begin{itemize}
\item ABG model:
\begin{multline}
\PL_\mathrm{ABG} (f_c, d)~ [\dB] =  10 \, \alpha \, \log_{10}({d}) + \beta + \\ 10 \, \gamma \, \log_{10}({f_c}) + \chi,
\label{PL_ABG}
\end{multline}
where $\alpha$, $\beta$, and $\gamma$ are the distance slope, intercept, and frequency slope, respectively. \revision{green!20}{(R2--1)}{}{For the ITU-R P.1411-13 ABG reference curves, we used the site-general LoS parameters $(\alpha,\beta,\gamma)=(2.07,\,31.23,\,2.06)$ and the NLoS parameters $(\alpha,\beta,\gamma)=(3.73,\,16.02,\,2.26)$~\cite{ITU_R_P_1411}.}
\end{itemize}
From the fitting results, it can be observed that in the LoS scenario, both the CI and FI models yield similar parameter values, such as the PLE ($n$ or $\alpha$) and the standard deviation ($\sigma_\chi$), for each frequency band. In contrast, under the NLoS scenario, the $\sigma_\chi$ values of the FI model for both the $154$~GHz and $300$~GHz bands are smaller than those of the CI model, suggesting that the FI model provides a better fit for NLoS conditions in this environment. Nonetheless, it is premature to draw definitive conclusions due to the limited number of samples.

From Fig.~\ref{LoS_PL}, \revision{green!20}{(R2--1)}{it can be observed that the measured PL of each frequency band has distinctly exhibited distance dependency and several dB improvement can be achieved over FSPL}{the measured PL at each frequency exhibits clear distance dependence and is slightly lower than the FSPL} due to multipath power in the LoS scenario. Furthermore, when comparing the measured PL with the FSPL in each band, the difference at $300$~GHz is slightly smaller than that at $154$~GHz. This indicates that the $300$~GHz band undergoes somewhat greater degradation, which is consistent with the cluster characteristics discussed in the previous section. On the other hand, the ITU-R model tends to overestimate the PL compared to the measured data, yielding values that exceed the FSPL. This discrepancy is likely because the ITU-R model is fitted using a vast dataset primarily from lower frequency bands, which may limit its prediction accuracy at sub-THz frequencies in this specific environment. 

In the NLoS scenario, the PLs are significantly higher than the FSPL, exhibiting a large shadowing distribution as evidenced by $\sigma_\chi$ of the fitting parameters shown in Fig.~\ref{NLoS_PL}. However, due to the limited sample size in this scenario, these parameters should be considered preliminary and indicative rather than definitive. The measured PL exhibits a milder slope than the ITU-R predictions. This deviation highlights the site-specific nature of the environmental geometry.

Relating our results to prior outdoor sub-THz measurements using the same CI model with a 1~m reference distance, our omnidirectional PLEs at $154$~GHz closely match the $142$~GHz UMi results reported in~\cite{Shakya2024} (LoS $n\approx1.96$, NLoS $n\approx2.92$). Overall, our near-free-space LoS slopes are consistent with street/campus measurements in the $140$--$371$~GHz range~\cite{Wilhelm2022,YangSC2023,WangSC2023}, while remaining differences (especially in NLoS) are mainly driven by site geometry and measurement/processing method.

\subsection{Large Scale Channel Parameters (LSPs)}
The PDP and PAP are calculated using \eqref{eq:PDP} and \eqref{eq:PAPx}, respectively. The obtained power profiles are then processed according to an \textit{acceptance criterion}. In particular, only those samples are retained that are either within $20$~dB of the peak sample in the power profiles \cite{ITU_R_P_1407} or within the cutoff level, whichever is higher. The temporal dispersion, RMS DS, is evaluated as the second moment of the PDP as follows
\begin{equation}
\DS = {\sqrt{\frac{\sum\limits_{\check{\tau}}{\check\tau^{2} \mathrm{PDP}(\check\tau)}}{\sum\limits_{\check{\tau}}{\mathrm{PDP}(\check\tau)}} - \left(\frac{\sum\limits_{\check{\tau}}{\check\tau \mathrm{PDP}(\check\tau)}}{\sum\limits_{\check{\tau}}{\mathrm{PDP}(\check\tau)}}\right)^2}} \label{eqn:ds}
\end{equation}
where each variable has the same meaning as has been explained earlier. 

Similarly, the corresponding PAP is processed according to the same \textit{acceptance criterion} before being used to evaluate the respective angular spreads (AS). Specifically, the AS of Departure (ASD) at the Tx and the AS of Arrival (ASA) at the Rx are calculated based on the definition in 3GPP TR 38.901 v18.0.0 \cite[Annex A]{3gppTR38901}, expressed as
\begin{equation}
\AS_{x} = {\sqrt{-2\ln{\left|\frac{\sum\limits_{\check{\phi}_x}{\exp{\bigl(\j\check\phi_x\bigr)}\cdot \mathrm{PAP}(\check\phi_x)}}{\sum\limits_{\check{\phi}_x}{\mathrm{PAP}(\check\phi_x)}}\right|}}},
\label{eqn:as}
\end{equation}
where the representation of each variable is as per the earlier definition. In the above equation, if $x=\T$, then $\AS_x$ refers to ASD; otherwise ($x=\R$), it refers to ASA. \revision{green!20}{(R2--3)}{}{We note that \eqref{eqn:as} follows the AS definition in 3GPP TR~38.901; however, instead of using cluster/subpath powers, we evaluate the AS directly from $\mathrm{PAP}(\check{\phi}_x)$.}

\begin{table}[t]
    \centering
    \setlength{\tabcolsep}{5pt} 
    \caption{LSP model parameters. Note that the 3GPP model is specified only up to $100$~GHz; thus, the values are extrapolated.}
    \label{LSPparameters}
    \begin{tabular}{cccrrrr}
        \toprule
        \multirow{3}{*}{Param.} & \multirow{3}{*}{Model}  & \multirow{3}{*}{}  
        & \multicolumn{2}{c}{LoS} & \multicolumn{2}{c}{NLoS$^*$} \\
        \cmidrule(lr){4-5}\cmidrule(lr){6-7}
        & & & $154$ & $300$ & $154$ & $300$ \\ 
        & & & GHz & GHz & GHz & GHz \\ 
        \midrule

        \multirow{4}{*}{\begin{tabular}{c}
            DS\\[-1pt]
            {\scriptsize $\log_{10}(\mathrm{DS}/1\mathrm{s})$}
        \end{tabular}}
        & \multirow{2}{*}{Meas.} 
            & $\mu$      & $-8.19$ & $-8.33$ & $-8.11$ & $-8.26$ \\
        &   & $\sigma$   & $0.31$  & $0.39$  & $0.89$  & $1.27$ \\
        \cmidrule(lr){2-7}
        & \multirow{2}{*}{3GPP} 
            & $\mu$      & $-7.67$ & $-7.73$ & $-7.36$ & $-7.42$ \\
        &   & $\sigma$   & $0.38$  & $0.38$  & $0.63$  & $0.68$ \\
        \midrule

        \multirow{4}{*}{\begin{tabular}{c}
            ASD \\[-1pt]
            {\scriptsize $\log_{10}(\mathrm{ASD}/1^{\circ})$}
        \end{tabular}}
        & \multirow{2}{*}{Meas.} 
            & $\mu$      & $1.19$  & $1.10$  & $1.47$  & $1.57$ \\
        &   & $\sigma$   & $0.23$  & $0.28$  & $0.47$  & $0.47$ \\
        \cmidrule(lr){2-7}
        & \multirow{2}{*}{3GPP} 
            & $\mu$      & $1.10$  & $1.09$  & $1.03$  & $0.96$ \\
        &   & $\sigma$   & $0.41$  & $0.41$  & $0.57$  & $0.60$ \\
        \midrule

        \multirow{4}{*}{\begin{tabular}{c}
            ASA \\[-1pt]
            {\scriptsize $\log_{10}(\mathrm{ASA}/1^{\circ})$}
        \end{tabular}}
        & \multirow{2}{*}{Meas.} 
            & $\mu$      & $1.01$  & $0.98$  & $1.00$  & $1.12$ \\
        &   & $\sigma$   & $0.18$  & $0.19$  & $0.56$  & $0.54$ \\
        \cmidrule(lr){2-7}
        & \multirow{2}{*}{3GPP} 
            & $\mu$      & $1.55$  & $1.53$  & $1.63$  & $1.61$ \\
        &   & $\sigma$   & $0.31$  & $0.31$  & $0.41$  & $0.42$ \\
        \midrule

        \multirow{4}{*}{$K$-factor [dB]}
        & \multirow{2}{*}{Meas.} 
            & $\mu$       & $7.01$  & $7.23$  & \NA & \NA \\
        &   & $\sigma$    & $4.50$  & $4.77$  & \NA & \NA \\
        \cmidrule(lr){2-7}
        & \multirow{2}{*}{3GPP} 
            & $\mu$       & \multicolumn{2}{c}{ $9$ } & \NA & \NA \\
        &   & $\sigma$    & \multicolumn{2}{c}{ $5$ } & \NA & \NA \\
        \bottomrule
        \multicolumn{7}{l}{\scriptsize $^{\mathrm{*}}$Preliminary values due to limited sample size.} \\        
    \end{tabular}
\end{table}

The power content in the LoS cluster compared to other significant MPCs is quantified as Rician $K$-factor and can be evaluated as
\begin{equation}
K = \frac{\sum_{\forall n_1} |\hat{\gamma}_{1, n_1}|^2}{\sum_{\forall k} \sum_{\forall n_k, k\ne 1} |\hat{\gamma}_{k,n_k}|^2}, \label{eqn:K}
\end{equation}
where $\hat{\gamma}_{k,n_k}$ represents the extracted complex path weight of the $n_k$-th MPC within the $k$-th cluster. $k=1$ denotes the LoS cluster.

The CDF plots of DS, ASD, ASA, and Rician $K$-factor for the considered scenarios are presented in Fig.~\ref{fig:LSP_dist}, and the corresponding LSP statistics (measured and 3GPP reference) are summarized in Table~\ref{LSPparameters}. 
\revision{green!20}{(R2--2)}{}{Since 3GPP TR~38.901 is specified for carrier frequencies up to $100$~GHz~\cite{3gppTR38901}, the 3GPP reference values at $154$~GHz and $300$~GHz were obtained by directly evaluating the TR~38.901 LSP-parameter equations at $f_\mathrm{c}=154$ and $300$ GHz. For parameters that are frequency-independent (constants) in TR~38.901, the same values as at $100$~GHz were used. This extrapolation is used only as a baseline reference for comparison.}

Based on the analysis, the key observations can be summarized as follows:
\begin{itemize}
\item The average DS is approximately $6.5$~ns at $154$~GHz and $4.7$~ns at $300$~GHz. Compared to the 3GPP extrapolation, the measured DS is generally smaller, indicating that the sub-THz channel in this environment is less dispersive in the delay domain than the standard model predicts.

\item In the LoS scenario, the average ASD is approximately \Ang{15.5} at $154$~GHz and \Ang{12.6} at $300$~GHz, while the average ASA is \Ang{10.2} and \Ang{9.5}, respectively. While the distributions remain statistically similar across the two frequencies, a slight reduction is observed at $300$~GHz. The ASD is consistently larger than the ASA, which can be attributed to the proximity of the Rx to the street walls (SW). Compared to the 3GPP model, the measured ASD shows reasonable agreement, whereas the measured ASA is notably smaller than the prediction. This discrepancy highlights the site-specific nature of the environment as well as the multipath sparsity characteristic of sub-THz bands.

\item The $K$-factors range from $0$ to $18$~dB, with an average of $7.01$~dB at $154$~GHz and $7.23$~dB at $300$~GHz (a slight
increase at $300$~GHz). No significant difference is observed between the distributions at the two frequencies. These results align well with the 3GPP model, which predicts a mean $K$-factor of $9$~dB, confirming that the dominance of the direct path remains consistent with lower-frequency extrapolations.

\end{itemize}

To statistically verify the frequency dependence, we conducted t-tests \cite{t-test} on the extracted log-valued LSPs. The results indicate no statistically significant differences ($p$-value, $p>0.05$) between $154$~GHz and $300$~GHz for any parameter. Specifically, \revision{green!20}{(R2--3)}{}{the Rician $K$-factor demonstrated significant similarity ($p=0.84$), while the dispersion parameters, DS ($p=0.22$), ASD ($p=0.27$), and ASA ($p=0.59$)}, also showed no significant shifts. However, their mean values did exhibit consistent physical trends: a slight increase in the $K$-factor and a slight decrease in dispersion parameters at $300$~GHz. These minor variations are attributed to increased interaction losses (e.g., rough surface scattering), which tend to suppress higher-order MPCs more severely than the LoS path. Consequently, this statistical evidence confirms that the channel's temporal and spatial dispersion is primarily governed by the site geometry rather than frequency-dependent scattering mechanisms.

The measurement configuration (BS height: $3$~m, and UE height: $1.2$~m) represents a low-height access link in a street-canyon environment (lamppost-installed BS to UE). The reported PL and LSPs are derived from the measurement distance ranges only (LoS: $10$--$80$~m, and NLoS: up to $40$~m). While the LoS dominance and multipath sparsity are expected to hold for similar street-canyon scenarios, the channel parameters may vary with street geometry/materials and, in particular, with Tx height due to changes in elevation geometry and the visibility/strength of wall and ground interactions.

\begin{figure*}[t]
\centering
\subfigure[Distribution of the differences between measured and calculated powers for the north and south walls, where Rx positions without values indicate signal absence. \label{fig:Diff_indvPG}]{\includegraphics[width=0.5\linewidth]{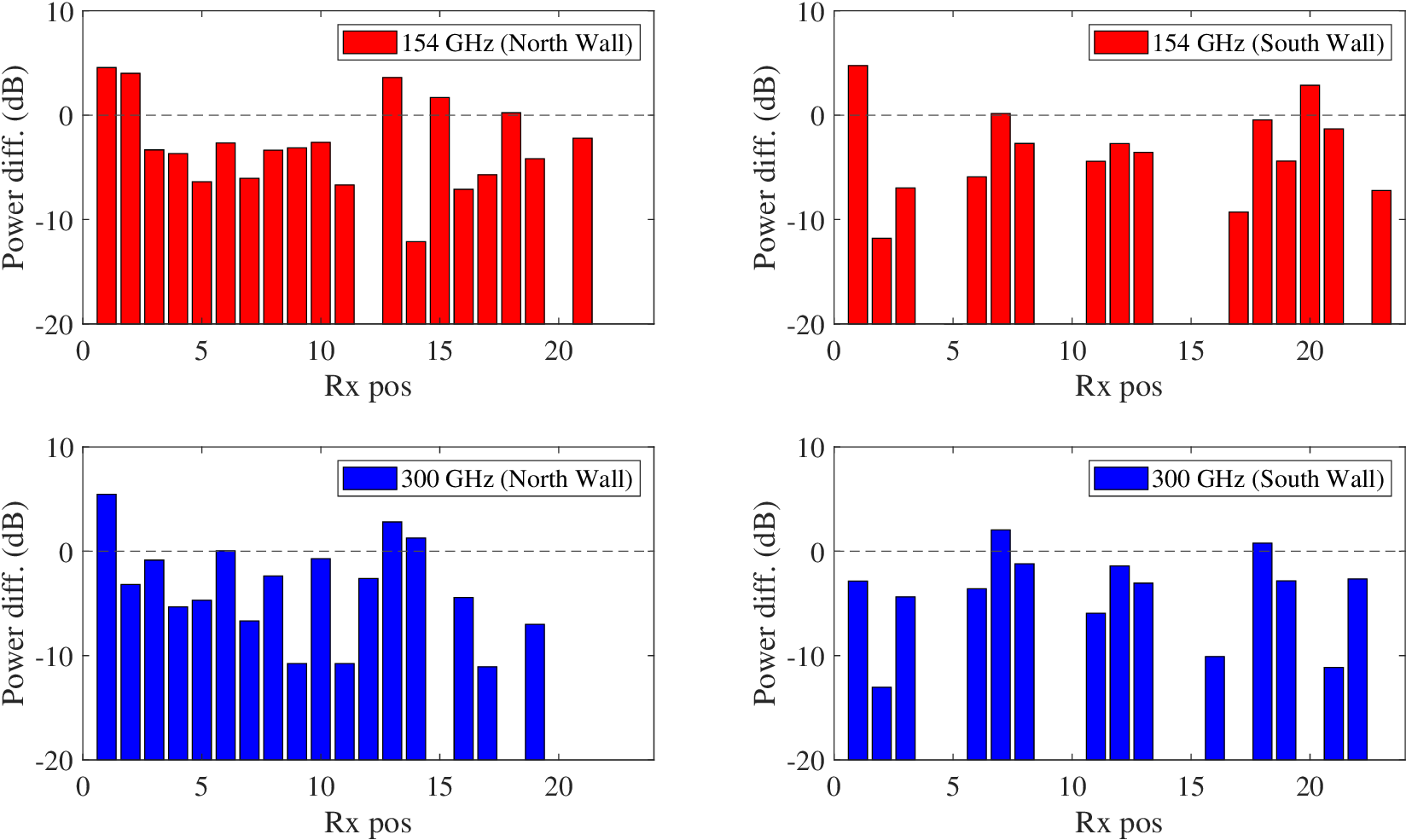}} \qquad
\subfigure[Distribution of the difference between measured and calculated powers.\label{fig:CDF_indvPG}]{\includegraphics[width=0.38\linewidth]{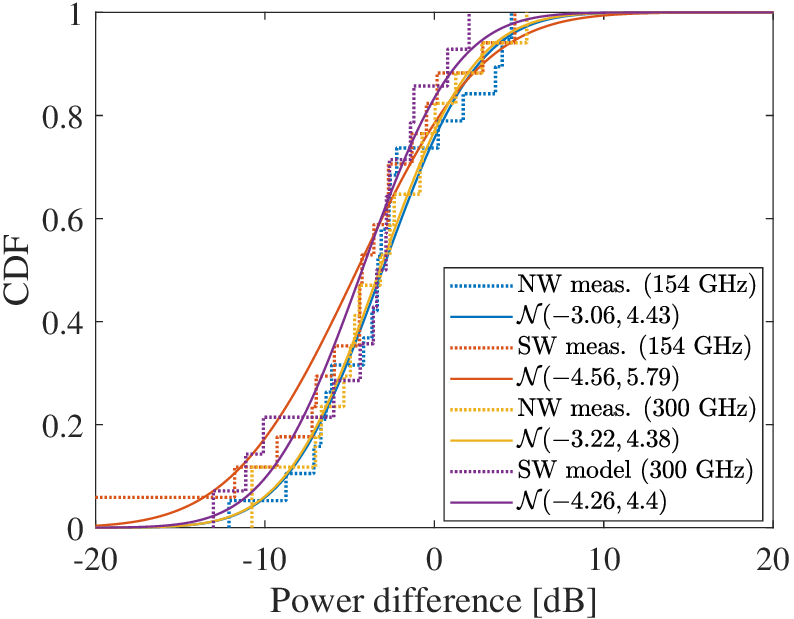}} 
\caption{Difference between measured and calculated powers of single-bounce wall-reflected paths. \label{fig:d_comp}}
\end{figure*}

\section{Channel modeling}
Based on the investigation of the propagation mechanisms, this section proposes a Q-D channel model inspired by those developed for corridor scenarios \cite{THzCorridor}. The discussion begins with the mathematical representation of the channel, followed by the modeling and generation of the deterministic and random components that constitute the channel. It should be noted that the channel model is developed only for the LoS scenario, due to the limited amount of available data.

\subsection{Q-D channel model}
Due to the short wavelength of sub-THz waves, specular reflection dominates and can be effectively predicted using deterministic methods such as ray tracing. Building on this characteristic as the deterministic component, a Q-D channel model \cite{Commag_Kim,Commag_Zemen} further accounts for non-negligible interactions arising from higher-order reflections and scattering caused by object shape and surface roughness, modeling them as random components. Namely, the propagation channel can be modeled as 
\begin{eqnarray}
h_c(\tau, \phi_\T, \phi_\R) =   h_c^\d(\tau, \phi_\T, \phi_\R) + h_c^\r(\tau, \phi_\T, \phi_\R),
\end{eqnarray}
where the deterministic components, $h_c^\d$, and random components, $h_c^\r$ are separately represented. Note that diffuse scattering is modeled as an additional contribution associated with both the deterministic and random components (i.e., as a local micro-cluster around each component).

As inferred in the previous section, the deterministic components are represented by the LoS path and the two dominant wall reflections from the NW and SW, and can be expressed as
\begin{multline}
h_c^{\mathrm{d}}(\tau, \phi_\T, \phi_\R) = \gamma_\los \delta(\tau_\los, \phi_{\T,\los}, \phi_{\R,\los}) + \\ \sum_{\forall l, l \in \{\nw, \sw\}} \chi_{l} \gamma_{l}^\d \delta(\tau_{l}^\d, \phi_{\T,l}^\d, \phi_{\R,l}^\d), \label{eq:deter}
\end{multline}
where `$\nw$' and `$\sw$' denote the single-bounce reflections by NW and SW, respectively. $\chi_l$ denotes a random variable that captures the randomness of the SB path existence and, in principle, its gain variations (e.g., due to surface roughness and associated diffuse scattering). In this work, however, $\chi_l$ is modeled using a two-state Markov process to represent the presence/absence of the SB path only, while a comprehensive stochastic modeling of the SB gain fluctuations (i.e., based on Fig.\ref{fig:CDF_indvPG}) is left for future work.

In addition, the random components are expressed as
\begin{eqnarray}
h_c^{\mathrm{r}}(\tau, \phi_\T, \phi_\R) = \sum \gamma_l^\r \delta (\tau_l^\r, \phi_{\T,l}^\r, \phi_{\R,l}^\r) \label{eq:rand}
\end{eqnarray}
where $\delta(\tau_l, \phi_{\T,l}, \phi_{\R,l}) \equiv \delta(\tau - \tau_l) \delta(\phi_\T - \phi_{\T, l}) \delta(\phi_\R - \phi_{\R, l})$ in both \eqref{eq:deter} and \eqref{eq:rand}.

\begin{table}
\centering
\caption{Estimated transition matrices for each frequency.}
\begin{tabular}{c|c}
\hline
Freq.  & Transition probability matrix \\ \hline
$154$~GHz & $\Vect{T}_{\text{NW}} = \begin{bmatrix} 0.333 & 0.667 \\ 0.158 & 0.842 \end{bmatrix}$, $\Vect{T}_{\text{SW}} = \begin{bmatrix} 0.500 & 0.500 \\ 0.186 & 0.813 \end{bmatrix}$ \\ \hline
$300$~GHz & $\Vect{T}_{\text{NW}} = \begin{bmatrix} 0.600 & 0.400 \\ 0.177 & 0.824 \end{bmatrix}$, $\Vect{T}_{\text{SW}} = \begin{bmatrix} 0.375 & 0.625 \\ 0.429 & 0.571 \end{bmatrix}$ \\ \hline
\end{tabular}
\label{tab:trans_mat}
\end{table}
\begin{figure}[t]
\centering
\includegraphics[width=0.8\linewidth]{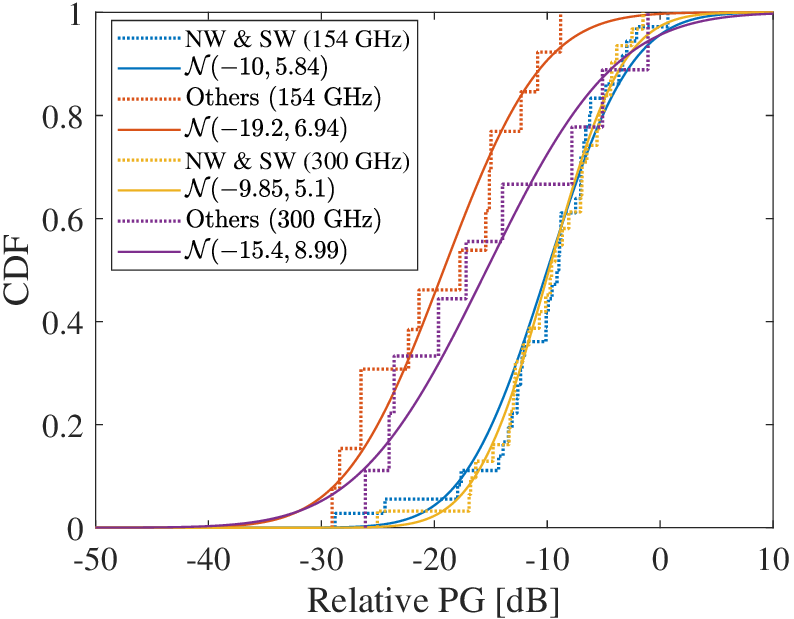}
\caption{Distributions of relative PG for deterministic and random components.\label{fig:CDF_relativePG}}
\end{figure}

\begin{figure*}[t]
\centering
\subfigure[CDF of interarrival time.\label{fig:Random_MPC_interarrival}]{\includegraphics[width=0.32\linewidth]{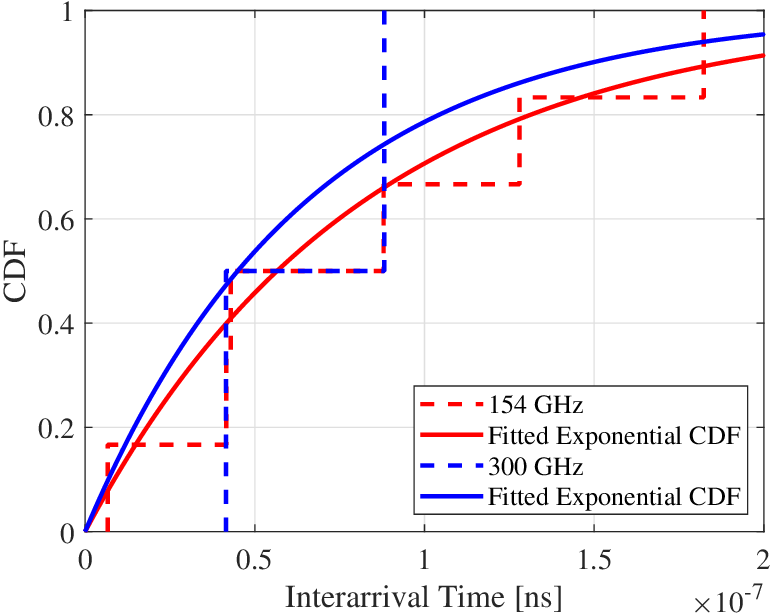}} 
\subfigure[Relative path gain to LoS.\label{fig:Random_MPC_PG}]{\includegraphics[width=0.32\linewidth]{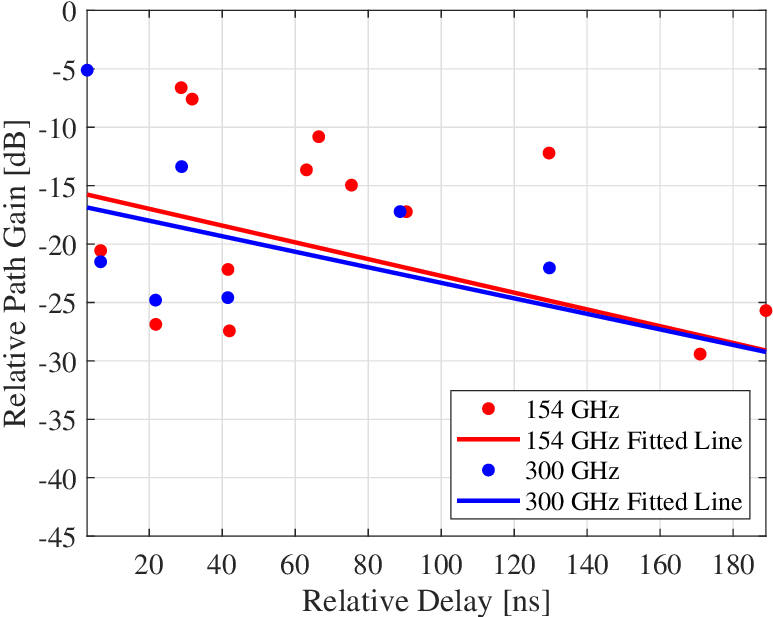}} 
\subfigure[Deviation to the model.\label{fig:Random_MPC_PG_diff}]{\includegraphics[width=0.32\linewidth]{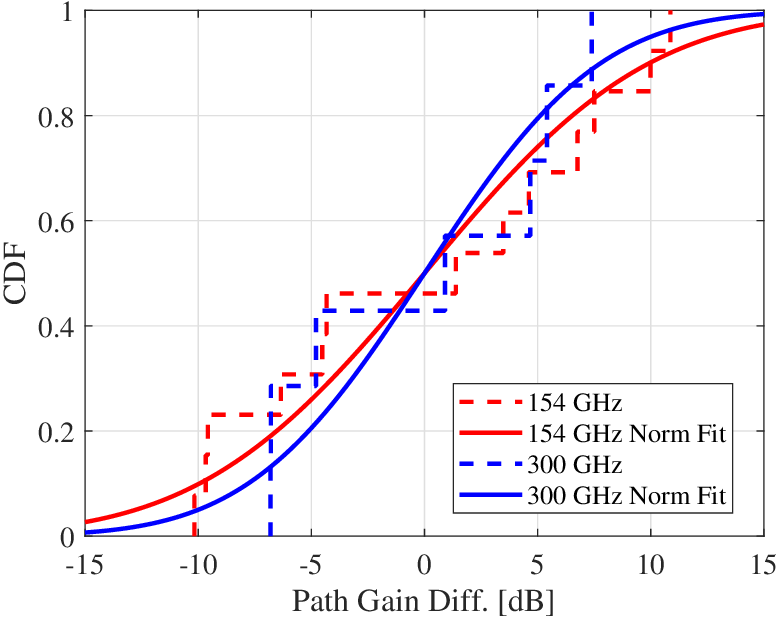}} 
\caption{Channel modeling for random MPCs. \label{fig:Random_MPC}}
\end{figure*}

\begin{table}[t]
\caption{Model parameters for random components}
  \label{tab:RandCompParams}
  \centering
  \begin{tabular}{c|c|c|c|c}
      \hline
      Freq.            & 1/$\lambda~[\mathrm{ns}]$ & $a$ & $b~[\dB]$ & $\sigma_\xi~[\dB]$ \\ \hline \hline
      $154$~GHz & $80.16$                    & $-0.07$              & $-15.55$            & $7.64$                 \\ \hline
      $300$~GHz & $64.82$                    & $-0.07$              & $-16.67$            & $6.87$                 \\
      \hline
  \end{tabular}
\end{table}

\subsection{Deterministic components}
When comparing the measurement results with ray-tracing simulations, it is often observed that certain reflected components appear randomly in the measurements and do not align well with the simulated results, thereby generating power differences, as illustrated Fig.~\ref{fig:d_comp}. A plausible cause of this trend may be the inhomogeneity in the wall shape and surface smoothness, along with the presence of various glass components such as windows and doors. 

Fig.~\ref{fig:Diff_indvPG} illustrates the distribution of the differences between the measured and calculated powers for the SB reflections from the NW and SW at both frequencies, where missing Rx positions indicate the absence of a signal. Fig.\ref{fig:CDF_indvPG} further presents the empirical CDFs together with their fitted models. To capture this probabilistic behavior induced by factors such as surface roughness, these models can be combined with a two-state Markov chain to represent signal presence or absence, as proposed in \cite{THzCorridor,Baum}. This property of the Markov chain is represented by the transition probability matrix $\Vect{T}= \begin{bmatrix} 1-p & p\\ q & 1-q \end{bmatrix}$ where $p$ is the probability of transitioning from `Absent' to `Present', and $q$ is the probability from `Present' to `Absent'. The modeling of path appearance is carried out separately for the SB reflections from the NW and SW at each frequency. Table~\ref{tab:trans_mat} summarizes the estimated transition probability matrices for each reflector. The steady-state probability of path absence, derived from the $(2,1)$ element of the limiting matrix $\Vect{T}_{\infty} = \lim_{n\rightarrow\infty}\Vect{T}^{n}$, increase from $154$~GHz to $300$~GHz. Specifically, the absence probability ($P_{\mathrm{absent}}=\frac{p}{p+q}$) rises from $0.1915$ to $0.3061$ for the NW reflection, and from $0.2727$ to $0.4068$ for the SW reflection. This trend is driven by greater scattering loss and diffuseness at $300$~GHz.

\subsection{Random components}
As described above, non-negligible random interactions should be incorporated into the channel model for a more accurate representation. Fig.~\ref{fig:CDF_relativePG} shows the distributions of relative PG for the deterministic and random components, where the SB reflections from the NW and SW are treated as deterministic components and all other components are treated as random. Note that this partition is adopted for the LoS scenario. For NLoS conditions, a dominant DB mechanism (e.g., SW$\rightarrow$NW in Fig.~\ref{fig:SP_transition_Del_NLoS}) may be incorporated as an additional deterministic component. Evidently, the random components make a significant contribution to the overall power. As explicated earlier, MPCs originating from non-predominant higher-order reflections and difficult-to-identify propagation paths are modeled as random components. 
\begin{figure}[t]
\centering
\includegraphics[width=0.9\linewidth]{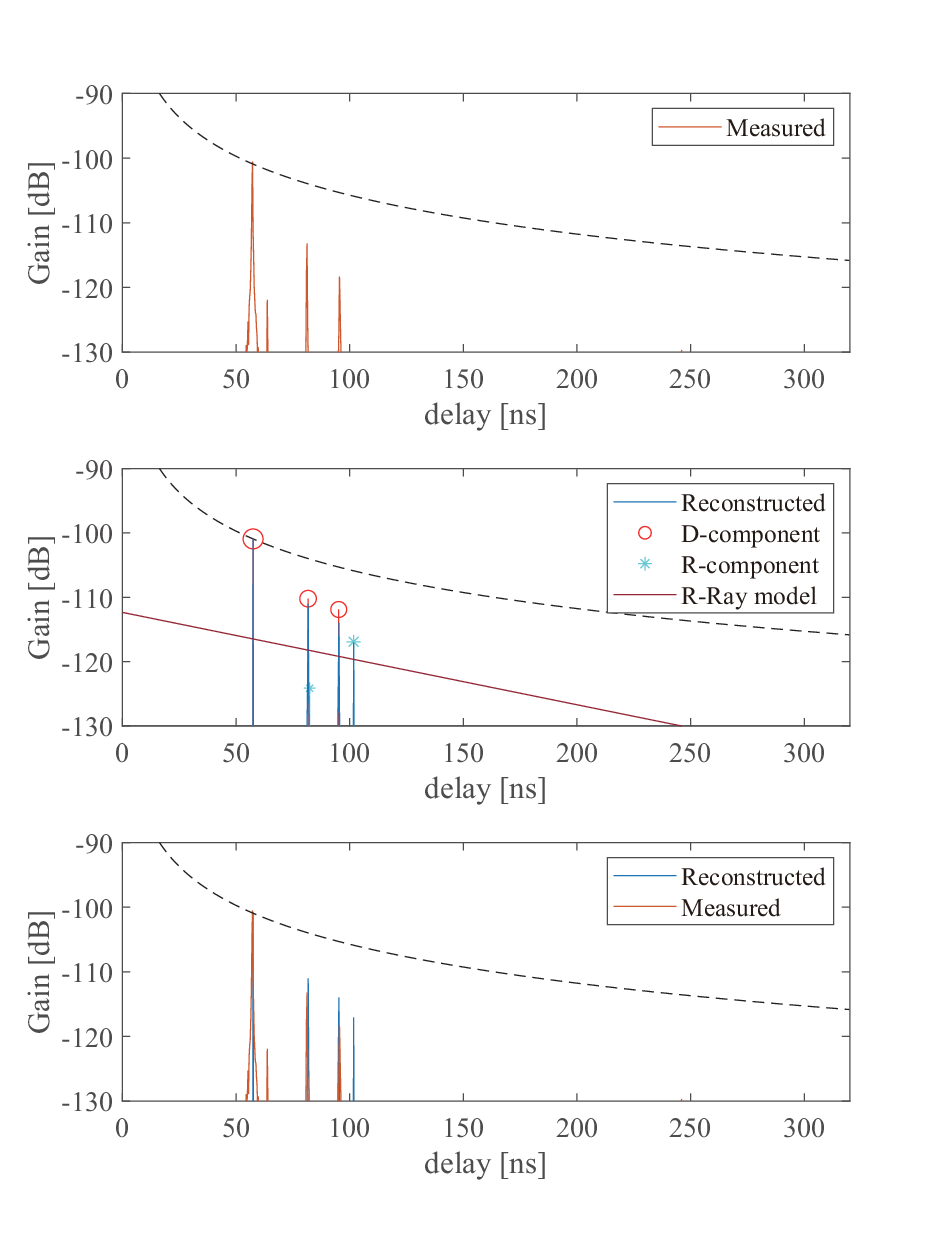}
\caption{Example of a CIR realization and comparison with a measured PDP in the delay domain ({\tt Rx3}), where dashed line denotes the free space path gain. Here, the D-component and R-component denote the deterministic and random components in the proposed Q-D model, respectively.}
\label{fig:PDPrealization}
\end{figure}

Following the procedure in \cite{THzCorridor}, Fig.~\ref{fig:Random_MPC_interarrival} presents the CDF of the interarrival times of these random components at both frequencies. The distribution of interarrival time is expected to follow an exponential distribution as 
\begin{equation}
  \label{eq:exp}
  F(\tau_k \mid \tau_{k-1}) = 1 - e^{-\lambda (\tau_k - \tau_{k-1})}, \quad \lambda > 0
\end{equation}
and has been fitted accordingly in Fig.~\ref{fig:Random_MPC_interarrival}. Here, ${\tau}_k$ represents the arrival time of the $k$th MPCs, and $\lambda$ is the reciprocal of the mean interarrival time (arrival rate). The fitting results presented in Table~\ref{tab:RandCompParams} show that the mean interarrival time at $154$~GHz is $80.16$~ns, which is higher than $64.82$~ns at $300$~GHz. It should be noted that these values are more than an order of magnitude larger than those observed in indoor corridor environments, owing to the outdoor setting.

In Fig.~\ref{fig:Random_MPC_PG}, the delay and power of the random components observed at each measurement location are normalized with respect to the delay and power of the LoS component at the same location. To model this characteristic, the path gain of the random component at delay $\tau$ and distance $d$ is as 
\begin{multline}
\label{eq:randomMPCs}
\PG^\mathrm{r}(\tau, d)~\mathrm{[dB]} = 
 a \cdot \frac{\Delta \tau(d)}{1~\mathrm{ns}} + b + \xi + \PG_\los(d)
\end{multline} where $\Delta \tau(d) = \tau - \tau_\los(d)$, $a$ represents the slope, and $b$ represents the intercept. Furthermore, $\tau_\los(d)$ and $\PG_\los(d)$ represent the delay and path gain of the LoS component at distance $d$, respectively. $\xi$ denotes a zero-mean Gaussian random variable with variance $\sigma_\xi^2$, representing shadow fading. The parameters $a$ and $b$ are estimated by fitting the plot in Fig.\ref{fig:Random_MPC_PG} using the maximum likelihood method \cite{Gustafson}, considering a noise level of $-30$ dB. The results yield $a = -0.07$ and $b = -15.55~\dB$ at $154$~GHz, and $a = -0.07$ and $b = -16.67~\dB$ at $300$~GHz, illustrating a similar trend at both frequencies. Furthermore, as shown in Fig.~\ref{fig:Random_MPC_PG_diff}, the path gain deviation to the model expressed in dB fits a normal distribution with a standard deviations of $\sigma_\xi = 7.64~\dB$ at $154$~GHz and $\sigma_\xi = 6.87~\dB$ at $300$~GHz, respectively. The obtained parameters for random cluster model are tabulated in Table \ref{tab:RandCompParams}. 

Fig.~\ref{fig:PDPrealization} presents an example of a CIR realization obtained through the proposed procedure, together with a comparison to the measured PDP in the delay domain ({\tt Rx3}) at $154$ GHz. The reconstructed PDP, derived using the same signal bandwidth and incorporating the radiation patterns of the antennas employed in the measurement, shows close agreement with the measured PDP. Because the model incorporates a stochastic component, the two PDPs cannot coincide exactly; nevertheless, the results demonstrate the validity of the proposed model.

\begin{figure}[t]
\centering
\includegraphics[width=\linewidth]{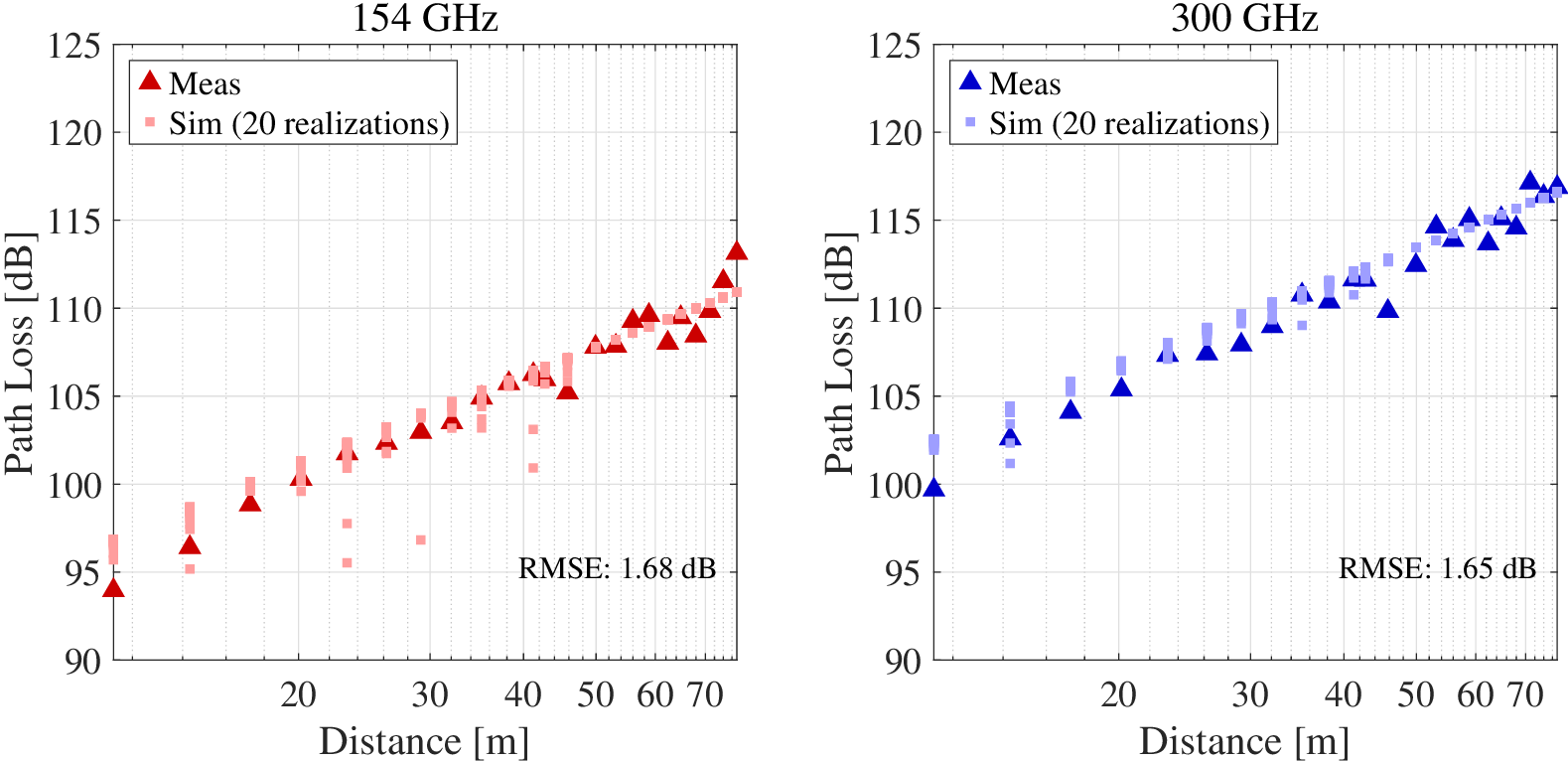}
\caption{Measured and synthesized path loss versus distance at (a) $154$~GHz and (b) $300$~GHz (multiple model realizations). The RMSE is indicated.}
\label{fig:Validation_PL}
\end{figure}
\begin{figure*}[t]
\centering
\subfigure[measured values and 20 model realizations.\label{Validation_DS_K}]{\includegraphics[width=0.48\linewidth]{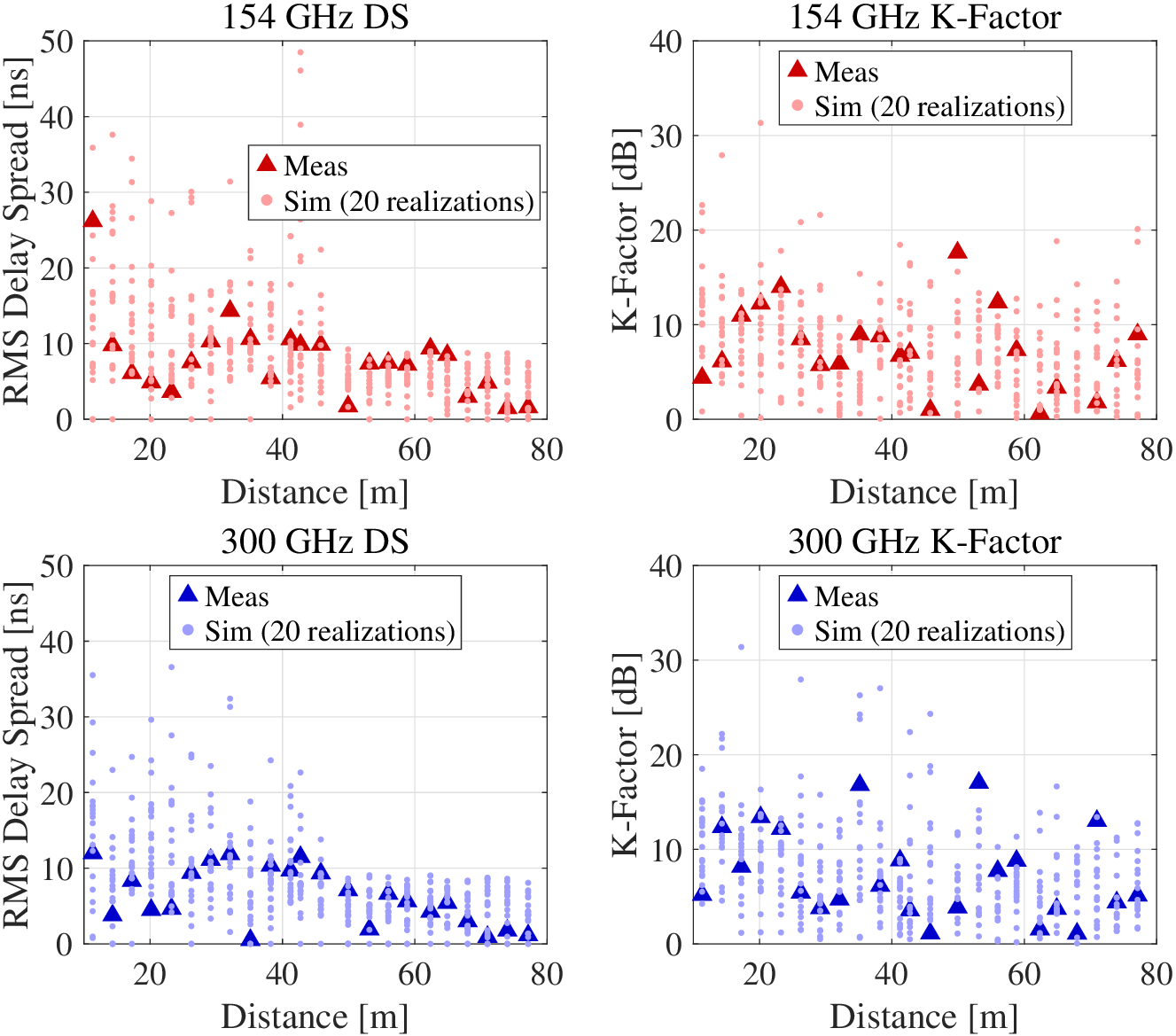}} 
    \subfigure[CDF with KS-test.\label{Validation_DS_K_CDF}]{\includegraphics[width=0.48\linewidth]{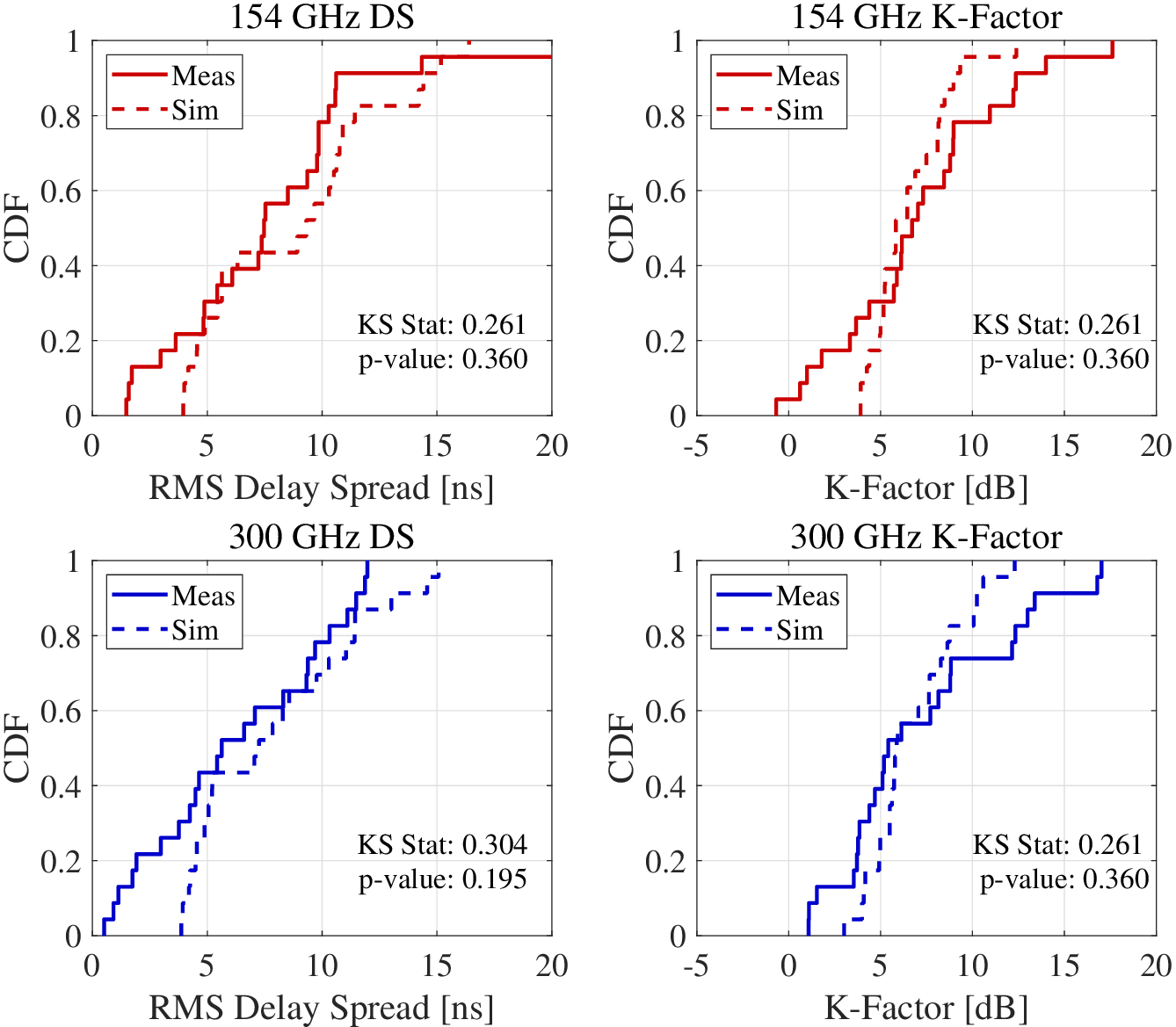}} 
\caption{Validation of DS and $K$-factor at $154$~GHz and $300$~GHz.}
\label{fig:Validation_DS_and_K}
\end{figure*}

\subsection{Quantitative Validation of the Proposed Model}
\label{subsec:quant_validation}
To assess the accuracy of the proposed model prediction, we quantitatively validate the generated channels against the measurement data.

\subsubsection{PL accuracy}
We first evaluate the PL prediction accuracy by comparing the measured and synthesized PL over all receiver locations. For each receiver position, the model generates multiple channel realizations to account for the stochastic components. Fig.~\ref{fig:Validation_PL} shows that the synthesized PL closely follows the measured distance-dependent trend at both $154$~GHz and $300$~GHz. The RMS error (RMSE) between measured and modeled PL remains small (approximately $2$~dB or less), confirming that the proposed model reliably captures the dominant behavior governed by site geometry.

\subsubsection{Statistical consistency of LSPs}
Next, we examine whether the model reproduces the statistical distributions of representative LSPs. Fig.~\ref{fig:Validation_DS_and_K} illustrates the validation results for DS and $K$-factor, displaying both the spatial variability (scatter plots) and the CDFs. We applied the two-sample Kolmogorov--Smirnov (KS) test to quantify the goodness of fit \cite{KStest}. As indicated, the KS statistics are consistently below the critical value, and the $p$-values exceed the significance level of $\alpha = 0.05$ in all cases. Consequently, the null hypothesis that the modeled and measured LSPs are drawn from the same distribution cannot be rejected.

The close overlap between the measured and modeled CDFs, combined with the favorable KS-test results, demonstrates that the proposed Q-D model accurately reproduces not only the average trends but also the variability of the dispersion characteristics. Importantly, the model effectively captures the frequency-dependent sparsity observed at $300$~GHz, reflecting the reduced multipath richness and weaker dispersion compared to the $154$~GHz band.

\section{Conclusion}
This paper presented an extensive D-D channel measurement campaign in a street-canyon environment at $154$~GHz and $300$~GHz, under both LoS and NLoS conditions, using an in-house-developed dual-band channel sounder. By applying joint clustering to merged datasets, we identified common multipath clusters and performed a comparative analysis of their frequency-dependent behaviors. The results show that the dominant propagation mechanisms are primarily the direct path and single-bounce reflections from the north and south walls. While statistical analyses of LSPs revealed no significant differences between the bands indicating geometric dominance, slightly greater interaction loss was observed at $300$~GHz. This is attributed to increased interaction losses (e.g., rough surface scattering), which tend to suppress higher-order MPCs more severely than the LoS path.

A Q-D channel model was proposed, integrating deterministic components, specifically the LoS and major reflections from the north and south walls, with random components representing higher-order or irregular multipath contributions. Furthermore, LSPs, including PL, DS, AS, and the Rician $K$-factor, were characterized. These contributions deepen the understanding of sub-THz propagation in urban street canyons and provide essential inputs for developing accurate, frequency-dependent channel models for future 6G systems.



\begin{thebibliography}{00}

\bibitem{WRC23Res255}
ITU, ``Resolution 255 (WRC-23): Studies on frequency-related matters for IMT identification in the frequency bands 102--109.5~GHz, 151.5--164~GHz, 167--174.8~GHz, 209--226~GHz and 252--275~GHz,'' \emph{Final Acts of the World Radiocommunication Conference 2023 (WRC-23)}, Dubai, UAE, 2023.

\bibitem{3gppTR38901}
3GPP, ``Study on channel model for frequencies from 0.5 to 100 GHz,'' 3rd Generation Partnership Project (3GPP), TR 38.901 version 18.0.0 Release 18, May 2024. [Online]. Available: \url{https://www.etsi.org/deliver/etsi_tr/138900_138999/138901/18.00.00_60/}

\bibitem{etsiTHz2024}
ETSI, ``Identification of use cases for THz communication systems,'' ETSI GR THz 001, Apr. 2024. [Online]. Available: \url{https://www.etsi.org/committee/1648-thz}

\bibitem{Commag_Rikkinen}
K. Rikkinen, P. Kyosti, M. E. Leinonen, M. Berg, and A. Parssinen, ``THz radio communication: Link budget analysis toward 6G,'' \textit{IEEE Commun. Mag.}, vol. 58, no. 11, pp. 22--27, Nov. 2020.

\bibitem{Commag_Kim}
M. Kim, J. Takada, M. Mao, C. Kang, X. Du, and A. Ghosh, ``THz channels for short-range mobile networks: Multipath channel behavior and human body shadowing effects,'' \textit{IEEE Commun. Mag.}, Early Access, doi: 10.1109/MCOM.001.2400694.

\bibitem{Wilhelm2022}
W. Keusgen, A. Schultze, M. Peter, and T. Eichler, ``Sub-THz channel measurements at 158~GHz and 300~GHz in a street canyon environment,'' in \textit{Proc. 2022 Joint European Conference on Networks and Communications \& 6G Summit (EuCNC/6G Summit)}, Grenoble, France, 2022, pp. 1--6.

\bibitem{Shakya2024}
D. Shakya, S. Ju, O. Kanhere, H. Poddar, Y. Xing, and T. S. Rappaport, ``Radio propagation measurements and statistical channel models for outdoor urban microcells in open squares and streets at 142, 73, and 28~GHz,'' \textit{IEEE Trans. Antennas Propag.}, vol. 72, no. 4, pp. 3580--3595, Apr. 2024.

\bibitem{YangSC2023}
W. Yang, Z. Yu, Y. Chen, M. Boban, T. Zugno, and J. Li, ``Channel measurements at 140 and 220~GHz in an outdoor street canyon environment,'' in \textit{Proc. 2023 IEEE Global Communications Conference (GLOBECOM)}, Kuala Lumpur, Malaysia, 2023, pp. 1477--1482.

\bibitem{WangSC2023}
Y. Wang, Y. Li, Y. Chen, Z. Yu, and C. Han, ``Terahertz channel measurement and analysis on a university campus street,'' in \textit{Proc. 2023 IEEE International Conference on Communications (ICC)}, Rome, Italy, 2023, pp. 2075--2080.

\bibitem{ITU_R_P_1411}
ITU-R, ``Propagation data and prediction methods for the planning of short-range outdoor radiocommunication systems and radio local area networks in the frequency range 300 MHz to 300 GHz,'' Recommendation ITU-R P.1411-13, Sept. 2025.

\bibitem{ITU_R_P_1238}
ITU-R, ``Propagation data and prediction methods for the planning of indoor radiocommunication systems and radio local area networks in the frequency range 300 MHz to 450 GHz,'' Recommendation ITU-R P.1238-13, Sept. 2025.

\bibitem{Commag_Zemen}
T. Zemen, J. Gomez-Ponce, A. Chandra, M. Walter, E. Aksoy, R. He, D. Matolak, M. Kim, J. Takada, S. Salous, R. Valenzuela, and A. F. Molisch, ``Site-specific radio channel representation for 5G and 6G,'' \textit{IEEE Commun. Mag.}, vol. 63, no. 6, pp. 106--113, Jun. 2025.

\bibitem{THzCorridor}
R. Takahashi, A. Ghosh, M. Mao, and M. Kim, ``Channel modeling and characterization of access, D2D, and backhaul links in a corridor environment at 300~GHz,'' \textit{IEEE Trans. Antennas Propag.}, vol. 73, no. 4, pp. 1954--1968, Apr. 2025.

\bibitem{Shakya_ICC}
D. Shakya \emph{et al.}, ``Urban Outdoor Propagation Measurements and Channel Models at 6.75 GHz FR1(C) and 16.95 GHz FR3 Upper Mid-Band Spectrum for 5G and 6G,'' 2025 IEEE International Conference on Communications (ICC), Montreal, QC, Canada, 2025, pp. 3291--3296

\bibitem{Kim_Access300GHz}
M. Kim, A. Ghosh, R. Takahashi, and K. Shibata, ``Indoor channel measurement at 300~GHz and comparison of signal propagation with 60~GHz,'' \textit{IEEE Access}, vol. 11, pp. 124040--124054, 2023.

\bibitem{Dupleich}
D. Dupleich, R. M\"{u}ller, and R. Thom\"{a}, ``Practical aspects on the noise floor estimation and cut-off margin in channel sounding applications,'' in \textit{Proc. 2021 15th European Conference on Antennas and Propagation (EuCAP)}, Dusseldorf, Germany, 2021, pp. 1--5.

\bibitem{Kim_CLEAN}
M. Kim \emph{et al.}, ``Millimeter-wave radio channel characterization using multi-dimensional sub-grid CLEAN algorithm,'' \textit{IEICE Trans. Commun.}, vol. E103-B, no. 7, pp. 767--779, Feb. 2020.

\bibitem{TVT_Tsukada}
H. Tsukada, N. Suzuki, B. Bag, R. Takahashi, and M. Kim, ``Millimeter-wave urban cellular channel characterization and recipe for high-precision site-specific channel simulation,'' \textit{IEEE Trans. Veh. Technol.}, vol. 74, no. 3, pp. 3598--3612, Mar. 2025.

\bibitem{concrete}
M. Urahashi, R. Okumura, K. Suizu, and A. Hirata, ``Complex permittivity evaluation of building materials at 200--500~GHz using THz-TDS,'' in \textit{Proc. Intl. Symp. Antennas Propag. (ISAP)}, Osaka, Japan, Jan. 2021, pp. 1--2.

\bibitem{ITU_R_P_1407}
ITU-R, ``Multipath propagation and parameterization of its characteristics,'' Recommendation ITU-R P.1407-8, Sept. 2021.

\revision{green!20}{(R2--1)}{}{
\bibitem{OmniSynth}
M. Kim and H. Yomoda, ``Synthesized-Isotropic Narrowband Channel Parameter Extraction from Angle-Resolved Wideband Channel Measurements,'' \textit{arXiv preprint} arXiv:2602.01646, Feb. 2026.
}

\bibitem{cimodel}
K. Haneda \emph{et al.}, ``5G 3GPP-like channel models for outdoor urban microcellular and macrocellular environments,'' in \textit{Proc. 2016 IEEE 83rd Vehicular Technology Conference (VTC Spring)}, Nanjing, China, 2016, pp. 1--7.

\bibitem{fimodel}
G. R. MacCartney \emph{et al.}, ``Path loss models for 5G millimeter wave propagation channels in urban microcells,'' in \textit{Proc. 2013 IEEE Global Communications Conference (GLOBECOM)}, Atlanta, GA, USA, 2013, pp. 3948--3953.

\bibitem{t-test}
B. L. Welch, On the comparison of several mean values: an alternative approach, Biometrika, vol. 38, no. 3-4, pp. 330--336, 1951.

\bibitem{Baum}
L. E. Baum, T. Petrie, G. Soules, and N.Weiss, ``A maximization techniques occurring in the statistical analysis of probability functions of Markov chain,'' \emph{Ann. Math. Stat.}, vol. 41, no. 1, pp. 164--171, 1970.

\bibitem{Gustafson}
C. Gustafson, D. Bolin, and F. Tufvesson, ``Modeling the cluster delay in mm-wave channels,'' in \textit{Proc. 8th Eur. Conf. Antennas Propag. (EuCAP)}, The Hague, Netherlands, Apr. 2014, pp. 1--5.

\bibitem{KStest}
F. Massey, ``The Kolmogorov-Smirnov test for goodness of fit,'' \emph{J. Amer. Stat. Assoc.}, vol. 46, no. 253, pp. 68--78, 1951.

\end{thebibliography}
\end{document}